# Amplified up-conversion of electromagnetic waves using time-varying metasurfaces

Fedor Kovalev,[1, 2,*] Stanislav Maslovski,[2] Abdelghafour Abraray,[2] and Ilya Shadrivov[1]

[1] ARC Centre of Excellence for Transformative Meta-Optical Systems (TMOS), Research School of Physics, The Australian National University, Canberra ACT 2601, Australia
[2] Instituto de Telecomunicações and Department of Electronics, Telecommunications and Informatics, University of Aveiro, Aveiro, 3810 – 193, Portugal

**ABSTRACT**. Time-varying metamaterials and photonic time crystals offer a powerful route to wave amplification through temporal modulation of material parameters. Here, we experimentally demonstrate amplified up-conversion of free-space electromagnetic waves in the microwave regime using a time-varying metasurface composed of varactor-loaded coupled split-ring resonators. The metasurface supports a cascaded parametric process in which temporal modulation enhances the resonant response near the signal frequency and subsequently enables efficient radiation at higher sum-frequency components. Direct measurements in degenerate and non-degenerate regimes show amplified up-conversion exceeding the single-step conversion estimate based on the Manley–Rowe frequency ratio. These results establish time-varying metasurfaces as an efficient platform for amplification, generation, and frequency conversion of electromagnetic waves in the microwave and terahertz bands, with potential extension to higher frequencies using ultrafast modulation mechanisms.

## I. INTRODUCTION

In time-varying or parametric metasurfaces, the physical parameters are modulated in time by an external energy source [1,2]. Time-varying media were first investigated more than fifty years ago [3–5] and have recently emerged as a highly promising area in metamaterial research [6,7], driven by the development of materials capable of ultrafast parameter modulation. The time dimension takes metasurface concept to a whole new level [6,8].

Time-varying metasurfaces (TVM) with physical parameters modulated in time go beyond the limits of passive electromagnetic media [9]. Such metasurfaces enable a range of non-trivial physical effects, including nonreciprocity [10–12], in which electromagnetic waves are transmitted differently when propagating in opposite directions. Recently, time-varying metasurfaces have enabled the realisation of photonic time crystals [13], a form of matter structured in time and analogous to conventional crystals structured in space, providing a new platform for the amplification of electromagnetic waves [14]. TVMs have also facilitated the demonstration of a temporal analogue of the Young's double-slit diffraction experiment [15]. Modulation of metasurface parameters over time gives rise to generation of new frequency harmonics when electromagnetic waves interact with such media, enabling frequency conversion [1,16]. Moreover, temporal interfaces recently demonstrated such exotic effects as time refraction and reflection [17–20], temporal counterparts of classical spatial physical processes.

Recent advances in TVMs have established them as a powerful platform for amplification, where temporal modulation and resonant enhancement allow electromagnetic waves to efficiently extract energy from the modulation process [21–26]. These results show that TVMs can not only convert and redistribute spectral content with high flexibility but also overcome the fundamental limitations of passive electromagnetic systems by providing controllable gain, paving the way for applications ranging from wireless communications to compact light sources. Amplification using TVMs has recently attracted significant attention owing to advances in ultrafast modulation techniques, which hold promise for its implementation at optical frequencies [19].

Parametric amplification can occur in degenerate and non-degenerate regimes. In the degenerate regime, the pump frequency is twice the signal frequency, $f_p = 2f_s$, and the phase relation between the signal and pump determines whether the signal is amplified or attenuated. In the non-degenerate regime, $f_p \neq 2f_s$, and amplification is accompanied by generation of an idler wave. In contrast to conventional guided-wave parametric amplifiers, the system investigated in this work performs amplified up-conversion of free-space waves using radiating time-varying meta-atoms. Importantly, the generated frequency components can lie above the pump frequency. This distinguishes the proposed metasurface from existing amplification platforms and time crystals that primarily amplify waves at the signal frequencies without converting the incident free-space wave to higher frequencies.

In this work, we experimentally demonstrate amplified up-conversion of *free-space electromagnetic waves* using TVMs. The proposed parametric meta-atoms achieve up-conversion gains exceeding 10 dB in the degenerate regime and 5 dB in the non-degenerate regime. The experimental results further confirm that, in certain cases, the up-conversion efficiency is not limited by the Manley–Rowe relations for the single-step conversion [27], owing to a cascaded amplification mechanism [28]. Notably, when the pump power approaches the self-excitation threshold, the radiated power at the sum frequency, which is approximately three times higher than the signal frequency, exceeds the incident signal power by more than a factor of three in both regimes.

These findings are directly relevant to microwave and sub-terahertz systems, where strong temporal modulation can be implemented using electronically controlled components. They point toward compact platforms for the generation, amplification, and frequency conversion of free-space electromagnetic waves. The demonstrated concept can be extended to higher-frequency regimes through the implementation of different ultrafast modulation mechanisms [19,29] and carefully engineered resonant responses [30].

## II. MODELING RESULTS

Figure 1 shows the proposed TVM, whose meta-atoms consist of dual split-ring resonators loaded with varactor diodes. The varactor capacitances are modulated by an external pump source through a biasing network at the frequency $f_p$. When an incident "signal" wave at $f_s$ interacts with the TVM, combination frequencies are generated, including the sum-frequency component $f_{sum1} = f_s + f_p$.

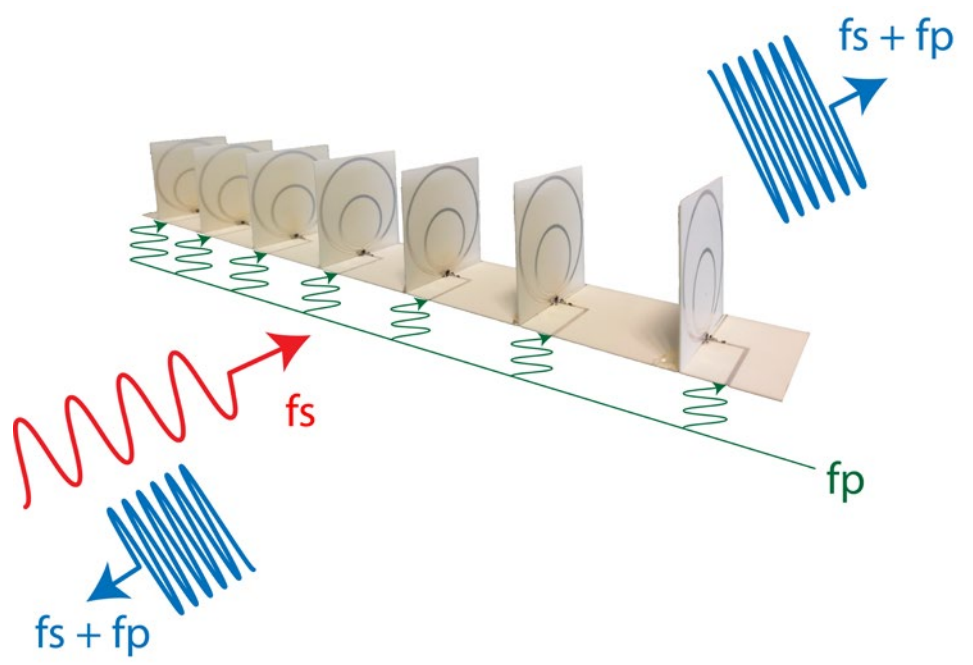

FIG 1. Schematic representation of amplified up-conversion by a TVM. An electromagnetic wave at $f_s$ is incident on the TVM, whose meta-atoms are modulated by a pump source at $f_p$, resulting in amplification and frequency conversion to the sum-frequency $f_s + f_p$.

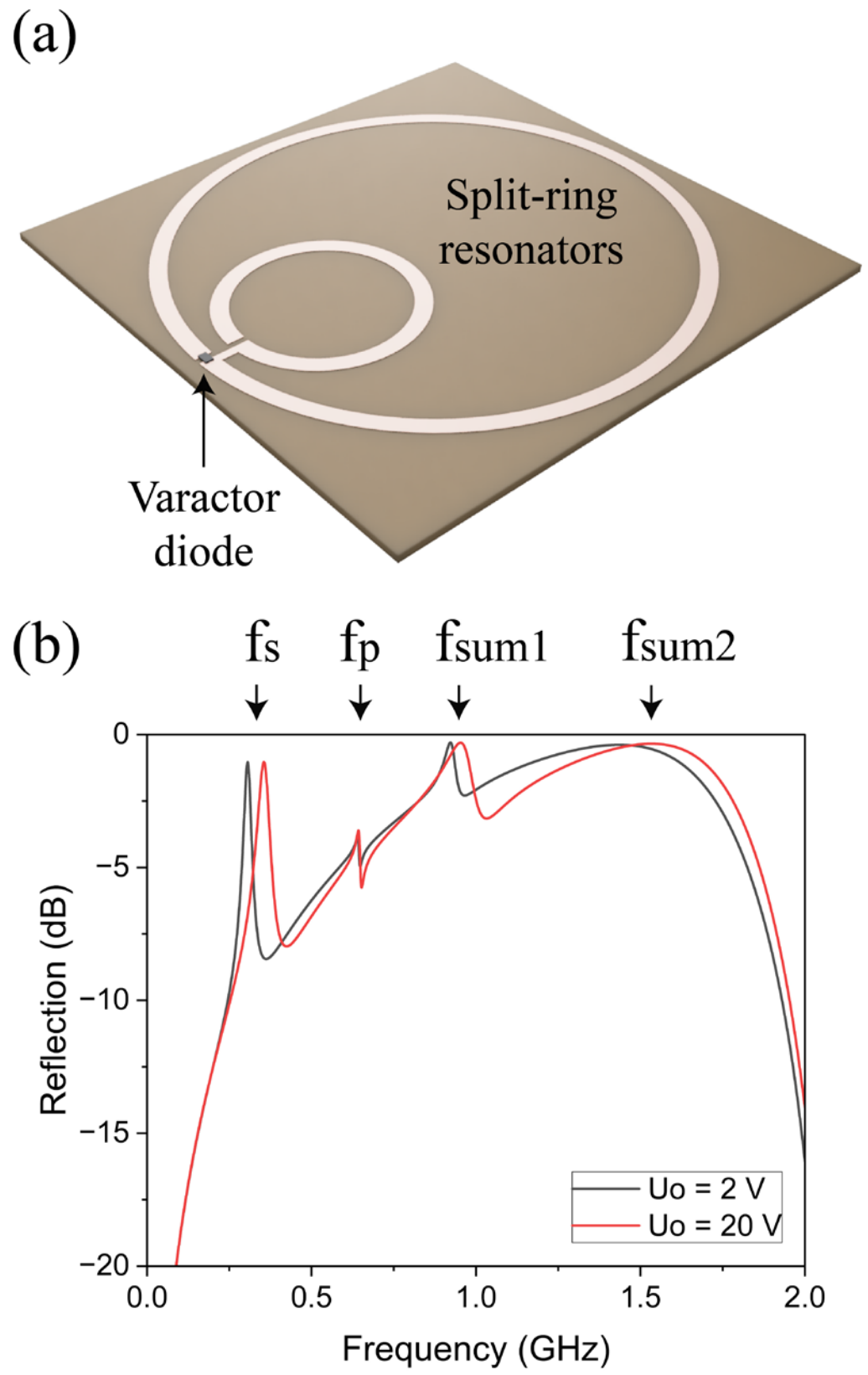

FIG 2. (a) Time-varying meta-atom consisting of two connected split-ring resonators with the embedded varactor diode. (b) Reflection characteristics of the metasurface for two applied dc voltage values ($U_0$).

The meta-atom consists of two connected split-ring resonators with the varactor diode (MGV100-20) embedded in the gap as illustrated in Figure 2(a). We first simulate the metasurface characteristics for two varactor diode capacitance values under the corresponding dc reverse voltage in CST Studio. See the meta-atom design, its equivalent circuit model and simulation details in the Supplemental Material. Figure 2(b) shows the reflection characteristics of the metasurface for two constant applied dc voltages ($U_0$). Two narrow resonances at $f_s$ and $f_{sum1}$ shift with variations in the applied voltage. The meta-atom geometry was optimized so that these resonances satisfy the approximate relation $f_{sum1} \approx 3f_s$. A broad resonance is also observed at $f_{sum2} \approx 5f_s$ in the spectrum, and its central frequency shifts with the applied dc bias voltage.

The reflection spectrum is used to identify resonances of the metasurface. The nonlinear generated power and its radiation direction are determined by the time-varying parametric process and are analyzed separately. The peak at $f_p$ does not originate from the meta-atom geometry itself, it appears in reflection due to the passband filter connected to the SRRs for protecting the ac pump source from the dc bias and currents at the signal and sum frequencies. We specifically choose $f_p \approx 2f_s$ and tune the resonances to the multiples of $f_s$ for utilizing the cascaded parametric amplification process allowing to overcome the limit imposed by Manley-Rowe relations for the single-step conversion [28].

*Contact author: fedor.kovalev@av.it.pt

For a single-step parametric up-conversion process, the Manley–Rowe relations provide a useful estimate for *the maximum power conversion ratio* between the generated sum-frequency component and the incident signal (for incommensurable $f_s$ and $f_p$):

$$K_P = \frac{P_{sum}}{P_s} = \frac{f_{sum}}{f_s}, \quad (1)$$

where $P_{sum}$ is the generated power at $f_{sum}$, $P_s$ is the power of the incident wave and $P_p$ is the power of the pump source. This estimate assumes that the signal and pump participate in a single conversion process and that the interacting powers are fully coupled into that process. In the present metasurface, however, the generated output results from a cascaded sequence of amplification and up-conversion processes supported by multiple resonances. We therefore use Eq. (1) as a reference for single-step conversion rather than as a fundamental bound on the total cascaded output.

In the degenerate regime, the pump frequency satisfies $f_p = 2f_s$, and the modulation provides phase-sensitive parametric amplification of the resonant response at the signal frequency. Because the metasurface also supports resonant responses near higher integer multiples of $f_s$, the amplified oscillation at $f_s$ can subsequently participate in further mixing with the pump and generate radiation at $f_{sum1}$ and $f_{sum2}$. In the non-degenerate regime, $f_p \neq 2f_s$, and the signal couples to an idler frequency $f_i = f_p - f_s$ through parametric modulation. The generated signal and idler components can then undergo further mixing with the pump, producing sum-frequency radiation near $f_{sum1}$, $f_{sum1i} = f_p + f_i$ and higher-order cascaded components: $f_{sum2} = f_s + 2f_p$ and $f_{sum2i} = f_i + 2f_p$.

Next, we discuss the results of the TVM modeling obtained in CST Studio and show that it is possible to overcome this limit for the single-step conversion when $f_{sum} \approx 3f_s$. See the simulation details in the Supplemental Material.

We modulate the varactor diodes integrated into the gaps of the dual split-ring resonators with the following pump voltage applied through microstrip lines and appropriate filtering circuits:

$$u_p(t) = U_0 + U_p cos(\omega_p t + \varphi_p), \quad (2)$$

where $U_0$ is the constant dc bias voltage, $U_p$ is the pump amplitude, $\omega_p = 2\pi f_p$ is the pump frequency, $\varphi_p$ is the pump phase.

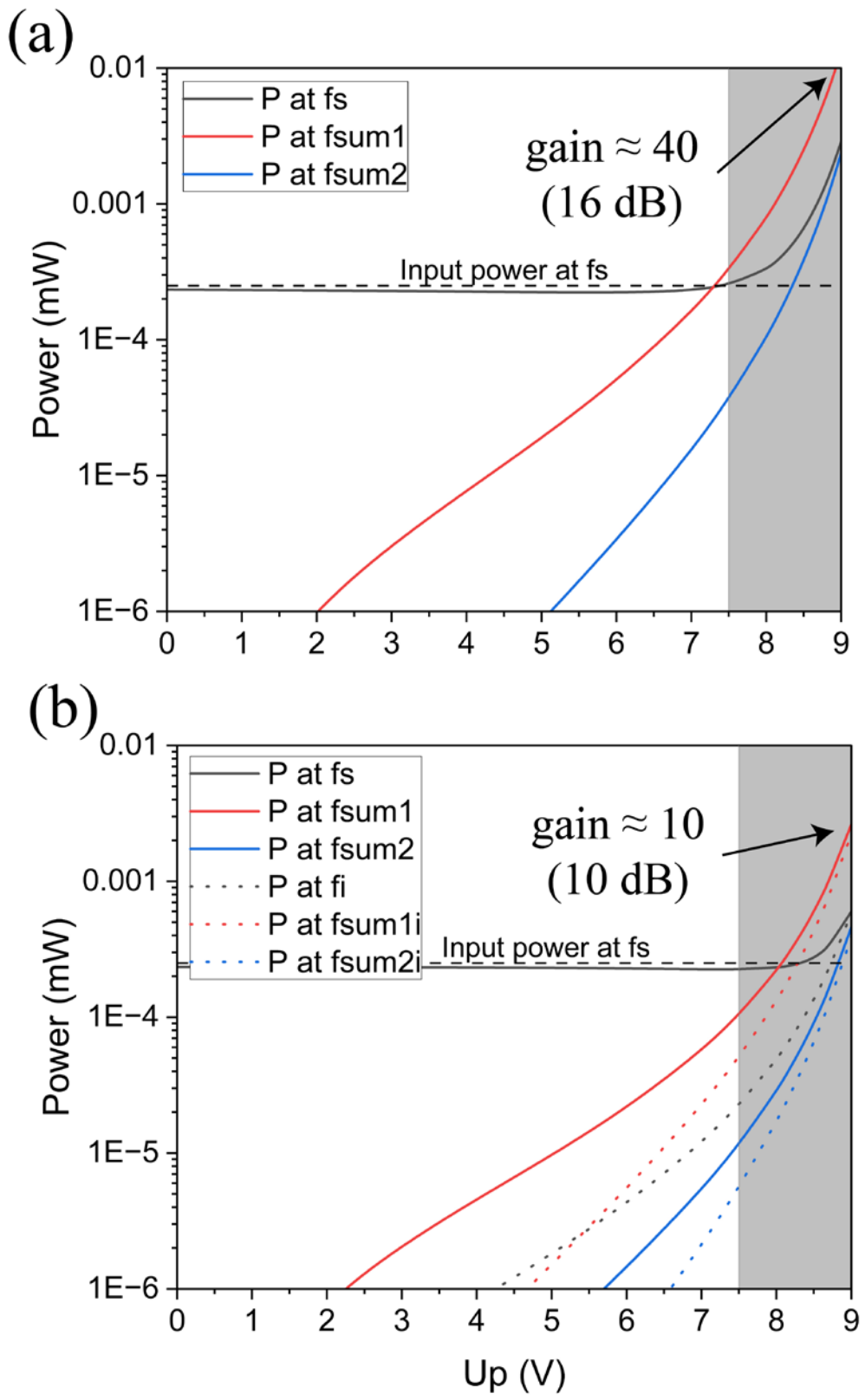

FIG 3. Calculated radiated power as a function of pump amplitude for $U_0 = 10$ V (logarithmic scale). (a) Degenerate regime under phase-locking conditions with $f_s =$ 320 MHz, $f_p =$ 640 MHz, $f_{sum1} =$ 960 MHz and $f_{sum2} =$ 1600 MHz; (b) Non-degenerate regime with $f_s =$ 320 MHz, $f_i =$ 319.9 MHz, $f_p =$ 639.9 MHz, $f_{sum1} =$ 959.9 MHz, $f_{sum1i} =$ 959.8 MHz, $f_{sum2} =$ 1599.8 MHz and $f_{sum2i} =$ 1599.7 MHz. The input power level of the "signal" wave is indicated by the horizontal dashed line. The shaded region corresponds to $U_p > 7.5$ V, where the varactor model becomes inaccurate according to its datasheet.

Figure 3(a) shows the dependence of the total radiated power at $f_s =$ 320 MHz, $f_{sum1} = 960$ MHz and $f_{sum2} = 1600$ MHz on the pump amplitude $U_p$ at $f_p =$ 640 MHz, for $U_0 = 10$ V under the condition of phase locking in the degenerate regime. The dashed horizontal line indicates the incident "signal" wave power at $f_s$, equal to 0.25 µW. The shaded region marks the range where the varactor model becomes inaccurate according to the manufacturer ($U_p >$ 7.5 V). Nevertheless, the selected varactor diodes operate in reverse bias from 0 to 22 V, with the latter value corresponding to breakdown. The results shown in Fig. 3 lie within this range.

*Contact author: fedor.kovalev@av.it.pt

Figure 3(a) shows that the total power radiated at $f_s$ increases rapidly once the pump amplitude exceeds 7 V, while remaining nearly constant below this threshold. The radiated power at both sum-frequency components also rises sharply with increasing pump amplitude. Figure 3(a) further demonstrates that *the overall radiated power at* $f_{sum1}$ *exceeds the input signal power*, indicating efficient up-conversion with gain. In particular, at a pump amplitude of $U_p = 9$ V, the overall radiated power at $f_{sum1}$ is more than forty times larger than the incident power at $f_s$, corresponding to a gain of approximately 16 dB. Moreover, under the same conditions, the radiated power at $f_{\mathrm{sum1}}$ exceeds the amplified power radiated at $f_s$ by more than a factor of four.

Figure 3(b) demonstrates that, in the non-degenerate regime, the overall radiated power at $f_{sum1}$ exceeds the input power at $f_s$ by more than an order of magnitude when the pump amplitude reaches 9 V and the pump frequency is set near the degenerate case ($f_p =$ 639.9 MHz). However, when the pump amplitude remains within the range of validity of the varactor model (below 7.5 V), the radiated power at both sum frequencies is lower than the incident power at $f_s$. The amplification and up-conversion characteristics in the non-degenerate regime closely resemble those obtained in the degenerate case. In the non-degenerate regime, the efficiency is reduced due to the simultaneous generation of power at the idler frequencies $f_i$, $f_{sum1i}$ and $f_{sum2i}$.

In the non-degenerate regime, the radiated sum-frequency power can be compared with the input signal power and with the single-step conversion estimate based on the Manley–Rowe frequency ratio. According to the Manley–Rowe relations, the maximum achievable power gain for the single-step conversion at $f_{\mathrm{sum1}}$ with respect to the incident signal power is limited by the frequency ratio $f_{\mathrm{sum1}}/f_s$ and in this case is close to three. The obtained output exceeding this reference indicates that the generated sum-frequency component is enhanced by cascaded amplification and up-conversion.

## III. EXPERIMENTAL RESULTS

Next, we describe the meta-atom design used in the experiments. Figure 4 presents the structure, which consists of two circuits fabricated on Rogers RO4350B substrates. The horizontal circuit board serves several functions: it supplies the pump power to the varactor diode, provides its dc bias and connects the split-ring resonators to the ground through a via. The back side of the substrate of the horizontal circuit board is copper-coated and functions as a ground plane. This arrangement provides the required isolation between the propagating electromagnetic wave and the pump circuit, ensuring that their influence on wave propagation is negligible. Each vertical board incorporates two connected copper split-ring resonators with an embedded MGV100-20 varactor diode and a blocking capacitor of 1 nF to protect the dc source connection to ground via the outer ring. The ac microstrip line provides pumping of the varactor diode through a passband filter consisting of inductor of 130 nH and capacitor of 0.1 pF, driven by a 50 Ω microwave generator. The dc microstrip line includes an inductor (RF choke) of 500 nH to block ac currents from reaching the dc source. See the details of the meta-atom design and geometry in the Supplemental Material.

We used a transverse electromagnetic (TEM) cell to emulate the response of the meta-atoms under plane-wave incidence. TEM cells offer a controlled environment for generating plane waves in a compact setting and detecting radiation from meta-atoms. Functioning as waveguides, they convert electrical signals into nearly homogeneous electromagnetic fields with a transverse mode distribution, closely resembling free-space conditions in modeling. Details of the TEM cell geometry, modeling of the time-varying meta-atoms in the TEM cell, and the experimental measurement scheme are provided in the Supplemental Material, along with a photograph of the experimental setup.

Figure 5(a) shows a photograph of the fabricated meta-atoms placed inside the TEM cell, along with biasing cables for connecting to a dc source and coaxial connectors for applying the ac pump modulation at $f_p$. Figure 5(b) presents the reflection characteristics of the meta-atoms within the TEM cell obtained from both simulations and experiments. Two resonances are observed at $f_s$ and $f_{sum1}$ (around 320 and 960 MHz, respectively, indicated by vertical lines for clarity).

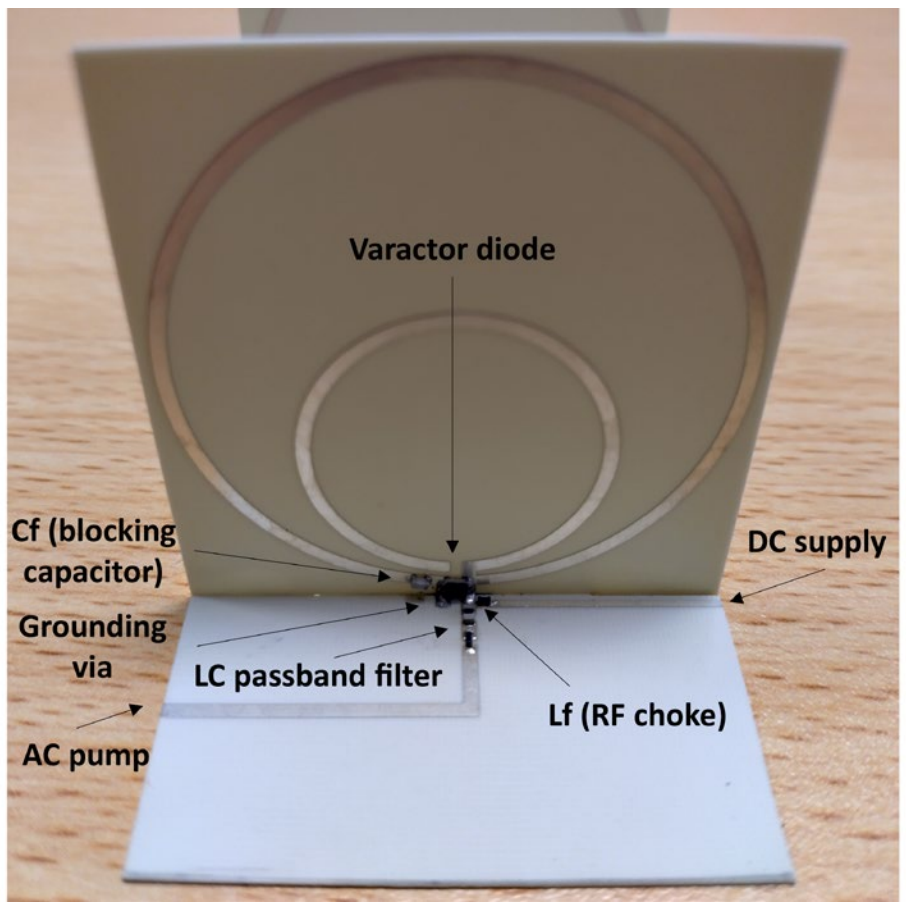

FIG 4. (a) Photograph of the fabricated meta-atom with soldered lumped elements.

*Contact author: fedor.kovalev@av.it.pt

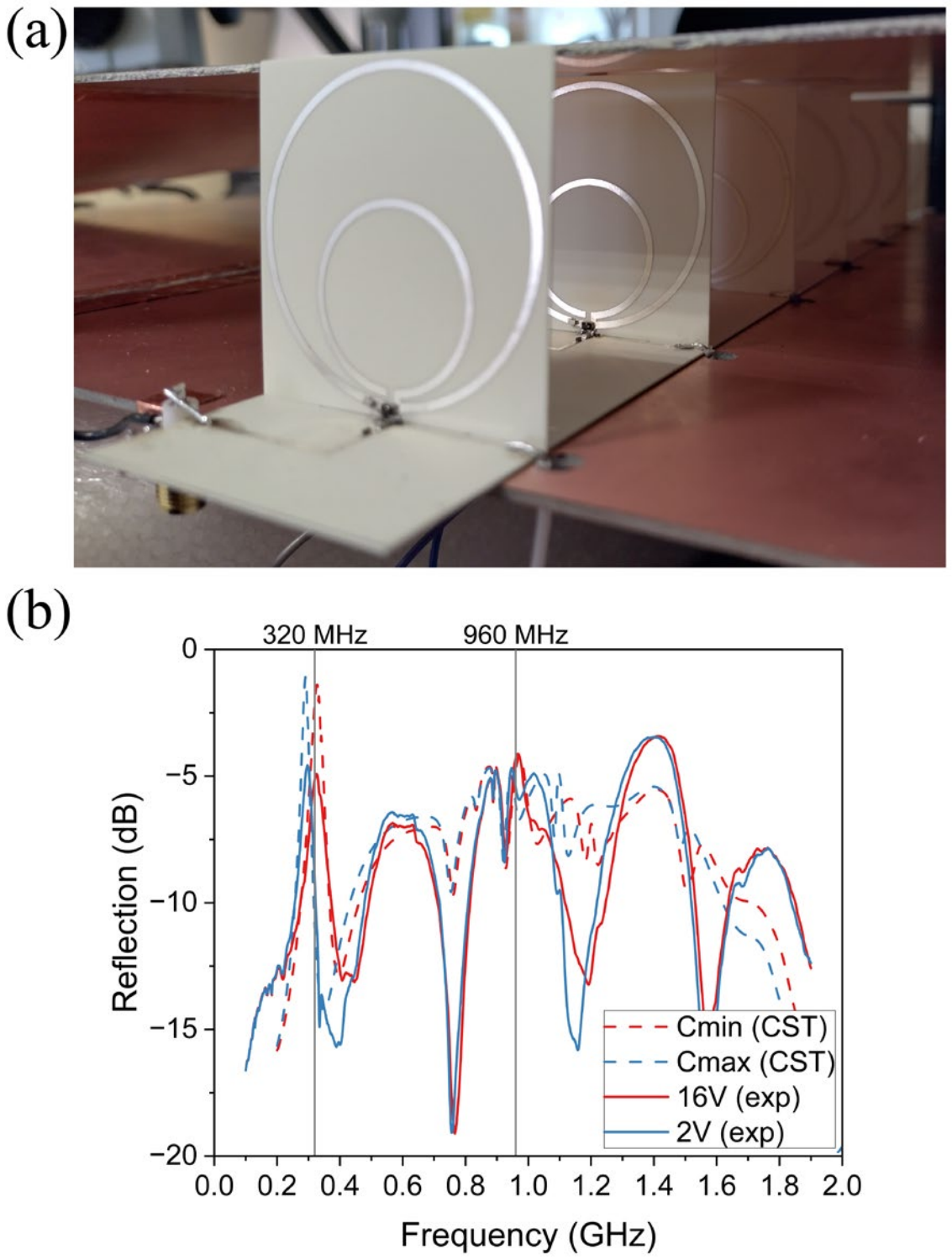

FIG 5. (a) Photograph of the fabricated meta-atoms inside the designed TEM cell. (b) Reflection characteristics of the meta-atoms in the TEM cell: simulations vs. experiments.

The resonances shift with changes in the capacitance of the varactor diodes in modeling and with the applied dc voltage in the experiments. Both simulation and experimental results show that the reflection level at $f_{sum1}$ is relatively low, primarily due to additional radiation and ohmic losses in the TEM cell at this frequency. The resonance at $f_s$ exhibits a reflection level that is more than 3 dB lower in experiments compared to simulations. This discrepancy is most likely caused by imperfect positioning of the meta-atoms within the TEM cell and fabrication imperfections.

Throughout the experimental analysis, the gain is defined as the ratio of the measured radiated power at the corresponding frequency to the input signal power at the generator output. The measured radiated power corresponds to the power collected by the TEM-cell measurement ports. Therefore, it is affected by losses in the resonators, biasing network, and imperfect coupling of the generated wave to the measured port. These results therefore confirm the possibility of amplified up-conversion in a radiating time-varying metasurface, while leaving room for further improvement through optimized metasurface design and more efficient collection of the generated radiation.

Significant up-conversion efficiency was experimentally achieved at a reduced dc bias voltage of $U_0 = 4$ V with pump power levels of up to 37 dBm at the output of the power amplifier, prior to splitting among the seven meta-atoms. This low-bias, high-pump regime enabled additional experimental investigation of the self-excitation threshold and parametric generation. However, this regime cannot be reliably reproduced in the numerical simulations, because the available SPICE model of the varactor diode is not specified for voltages below 2 V, whereas strong pump modulation can drive the diode outside this validity range during part of the modulation cycle. Therefore, the simulations are used to capture the main parametric mixing trends, while quantitative discrepancies at high pump powers may arise from additional nonlinear effects in the real circuit that are not included in the available varactor model.

Figure 6 shows the dependence of the overall radiated power at $f_s =$ 316 MHz, $f_{sum1} = 948$ MHz and $f_{sum2} =$ 1580 MHz on the pump power at $f_p =$ 632 MHz (at the output of the power amplifier) for $U_0 =$ 4V under phase-locking in the degenerate regime (logarithmic scale, dBm): (a) illustrates the overall response; (b) highlights the distinction between parametric amplification and generation; (c) and (d) present results in the reflection and transmission directions, respectively, relative to the propagation of the incident wave at $f_s$. The self-excitation threshold, at 35.7 dBm, is marked by a black vertical line at the edge of the shaded area. The input power at $f_s$ (−20 dBm at the generator output) is indicated by a horizontal dashed line. Remarkably, the system shows hysteresis of the amplification and generation. In Figure 6, the solid lines correspond to measurements where the pump power was increased from 30 to 37 dBm, while the dotted lines show the reverse sweep from 37 to 30 dBm.

Figure 6(a) demonstrates that the overall radiated power at $f_s$ is approximately −26 dBm at a pump level of 30 dBm. It increases slightly with pump power, reaching −25 dBm at a pump level of 35 dBm, before rising sharply up to −17 dBm. This corresponds to a gain of ~3 dB relative to the input at $f_s$. The growth saturates around 35.5 dBm, close to the self-excitation threshold, and reaches −16 dBm at 37 dBm. A marked rectangle in Figure 6(a) highlights the section shown in Figure 6(b) for comparison with the generated power when no incident wave at $f_s$ was present and the pump power exceeded the self-excitation threshold.

Figure 6(b) demonstrates the hysteresis loop: when decreasing the pump power from 37 dBm, the dotted curve follows the solid curve almost until the self-excitation threshold, after which the loop becomes evident. The radiated powers at $f_{sum1}$ and $f_{sum2}$ exhibit rapid growth with pump power. At $f_{sum1}$, the power surpasses the input level at 35.25 dBm and approaches −8 dBm near the self-excitation threshold, corresponding to 12 dB amplification relative to the input at $f_s$. Notably, the radiated power at $f_{sum1}$ is approximately 9 dB higher than that at $f_s$. At the self-excitation threshold, the powers at $f_{sum1}$ and $f_{sum2}$ reach −7 dBm and −20 dBm, respectively. Beyond threshold, the amplified up-conversion at $f_{sum1}$ saturates, while the radiated power at $f_{sum2}$ continues to increase slightly. The dotted lines reveal hysteresis in both sum-frequency powers, consistent with the behaviour at $f_s$.

*Contact author: fedor.kovalev@av.it.pt

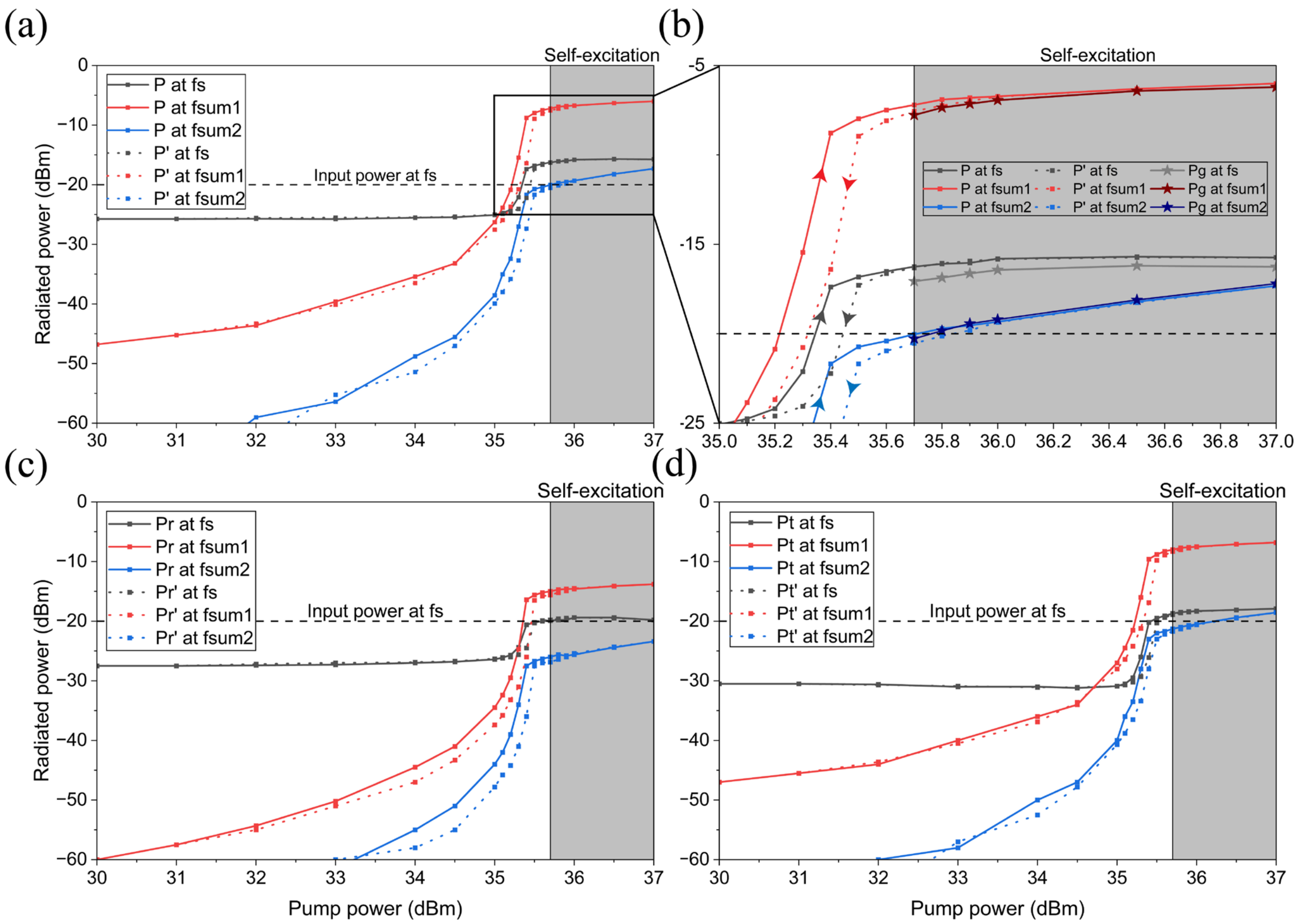

FIG 6. Dependence of the radiated power at $f_s$ = 316 MHz, $f_{sum1}$ = 948 MHz and $f_{sum2}$ = 1580 MHz on the pump power at $f_p$ = 632 MHz for $U_0$ = 4 V under phase-locking in the degenerate regime (logarithmic scale, dBm). (a) Overall radiated power; (b) same as (a) but zoomed into the region of the boundary between maximum amplification and self-generation; (c, d) power radiated in the directions of reflection and transmission, respectively, relative to the incident wave at $f_s$. Solid lines correspond to increasing pump power, while dotted lines correspond to decreasing pump power. The self-excitation threshold is indicated by a black vertical line at the edge of the shaded area, and the input power at $f_s$ is shown as a horizontal dashed line.

Figure 6(b) highlights the hard self-excitation process, characterized by an abrupt onset of radiation ($P_g$) at $f_s$, $f_{sum1}$ and $f_{sum2}$ when the pump exceeds 35.7 dBm. Interestingly, Figure 6(b) shows that in the absence of an incident signal at frequency $f_s$, the generated power $P_g$ above the self-excitation threshold follows a similar dependence on the pump power, with the power radiated at $f_{sum1}$ being an order of magnitude higher than the power radiated at $f_s$.

Figures 6(c) and 6(d) present the directional dependence of the amplification and generation processes. In reflection, the power at $f_s$ grows with pump power, mirroring the overall trend. In transmission, however, the power at $f_s$ initially decreases by 1 dB before increasing rapidly above 35 dBm. The sum-frequency behaviour is similar to that discussed above. Most of the power at $f_{sum1}$ and $f_{sum2}$ is radiated in transmission: at 35.5 dBm, the power levels in reflection are −16 dBm and −27 dBm, compared to −9 dBm and −22 dBm in transmission.

Figure 7 shows the dependence of the overall radiated power at $f_s$ = 316 MHz, $f_{sum1}$ = 947.9 MHz and $f_{sum2}$ = 1579.8 MHz on the pump power at $f_p$ = 631.9 MHz for $U_0$ = 4 V in the non-degenerate regime (logarithmic scale, dBm).

*Contact author: fedor.kovalev@av.it.pt

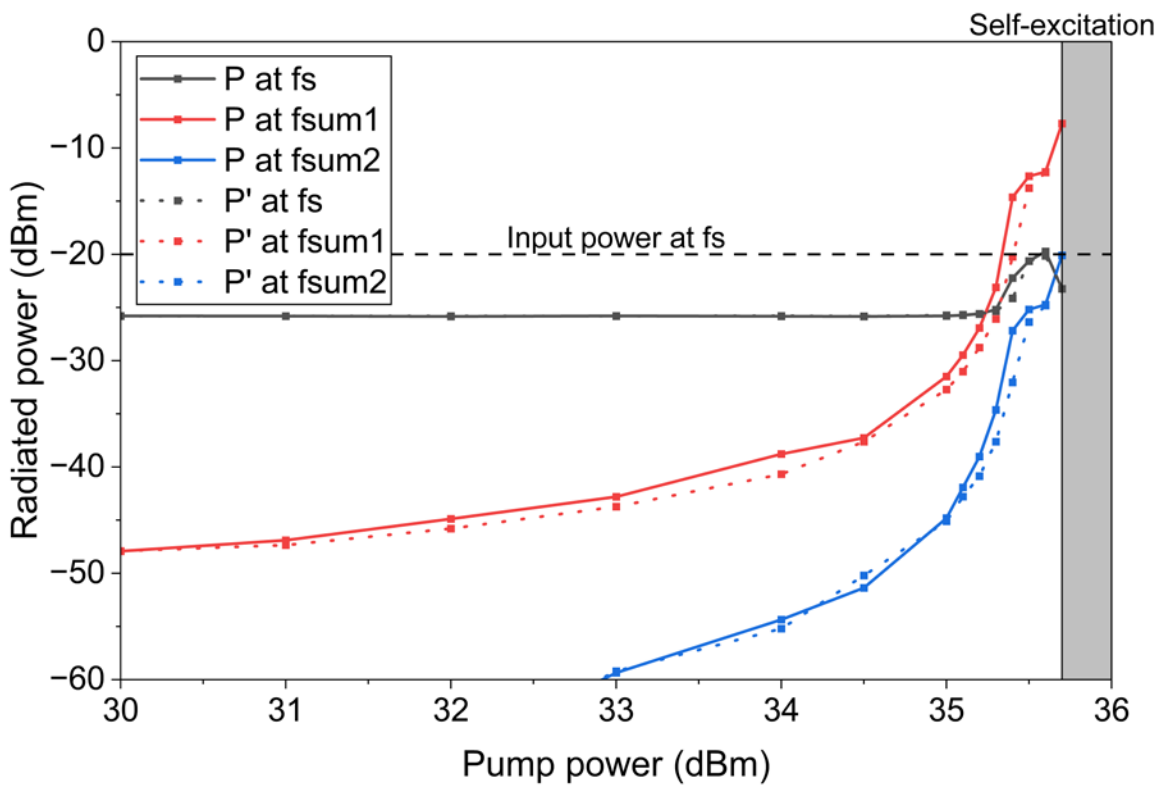

FIG 7. Overall radiated power at $f_s = 316$ MHz, $f_{sum1} = 947.9$ MHz and $f_{sum2} = 1579.8$ MHz as a function of the pump power at $f_p = 631.9$ MHz for $U_0 = 4$ V in the non-degenerate regime (logarithmic scale, dBm). Solid lines indicate increasing pump power, while dotted lines indicate decreasing pump power. The vertical line marks the self-excitation threshold, and the horizontal dashed line denotes the input power level at $f_s$.

The radiated power at $f_s$ remains nearly constant (–26 dBm) until the pump power reaches 35 dBm. Beyond this point, the power grows rapidly, reaching a maximum of about –20 dBm just below the self-excitation threshold (35.7 dBm), where it drops abruptly by 3 dB. This sharp decrease indicates that part of the pump power is diverted to generate an additional component at $f_p/2 = 315.95$ MHz due to the self-excitation process. Amplification at $f_s$ occurs primarily in transmission.

Meanwhile, the powers radiated at $f_{sum1}$ and $f_{sum2}$ increase rapidly with pump power up to about 35.5 dBm, after which they begin to saturate. At the instability threshold, both show an abrupt increase of approximately 4 dB. Remarkably, the overall radiated powers at $f_{sum1}$ and $f_{sum2}$ reach –8 dBm and –20 dBm, respectively, at the self-excitation threshold. These values are comparable to those observed in the degenerate regime. Although contributions are present in both reflection and transmission, the transmitted powers at both sum-frequencies dominate. At a pump power of 35.5 dBm, the overall radiated powers are about –21 dBm at $f_s$, –13 dBm at $f_{sum1}$, and –25 dBm at $f_{sum2}$. This indicates that the radiated power at $f_{sum1}$ exceeds that at $f_s$ by more than 7 dB in the non-degenerate regime. At this pump level, the radiated powers at $f_{sum1}$ and $f_{sum2}$ are –20.2 dBm and –31 dBm in reflection, and –13.5 dBm and –26.5 dBm in transmission, respectively.

Figure 8 presents the dependence of the radiated power at $f_s = 316$ MHz, $f_{sum1}$ and $f_{sum2}$ on the pump frequency for $U_0 = 4$ V and a pump power of 35.5 dBm in the non-degenerate regime. When the pump frequency is close to twice the signal frequency ($f_p \approx 2f_s$), the radiated power at $f_s$, $f_{sum1}$ and $f_{sum2}$ exhibits periodic temporal variations. These arise from beating between the signal and idler waves, with the beat frequency equal to their frequency difference. This beating modulates the amplitude of the radiated waves, producing fluctuations in their measured power at the beat frequency. The beating range is shown as the shaded areas in Figure 8. Away from this condition, the up-conversion efficiency decreases as the pump frequency deviates from the degenerate value. As shown in Figure 8(b), the transmitted power at $f_{sum1}$ exceeds the input power level within the frequency range 631.65–632.35 MHz, corresponding to an amplified up-conversion bandwidth of less than 1 MHz.

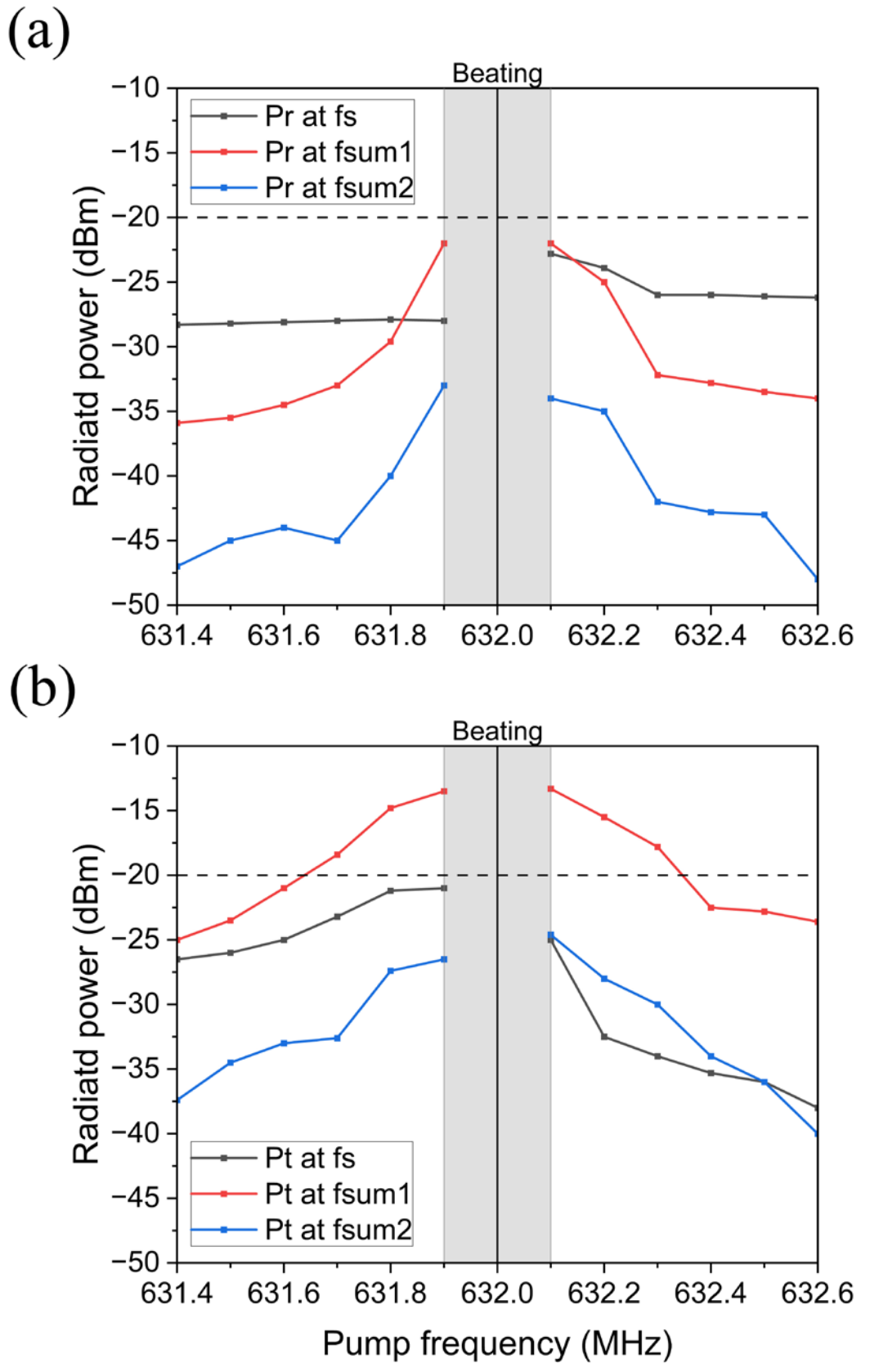

FIG 8. (a, b) Radiated power at $f_s = 316$ MHz, $f_{sum1}$ and $f_{sum2}$ as a function of pump frequency for $U_0 = 4$ V and a pump power of 35.5 dBm in the non-degenerate regime (logarithmic scale, dBm). The degenerate condition ($f_p = 632$ MHz) is indicated by a vertical line, while the input power at $f_s$ (–20 dBm) is shown as a horizontal dashed line. The shaded region near the degenerate regime marks the frequency range where beating occurs in the non-degenerate regime.

*Contact author: fedor.kovalev@av.it.pt

Furthermore, the experiments revealed a linear dependence of the phase of the radiated waves at $f_s$, $f_{sum1}$, and $f_{sum2}$ on the pump phase in the non-degenerate regime, which can be used for beam steering through the appropriate spatiotemporal modulation of the metasurface parameters.

## IV. DISCUSSION

We experimentally demonstrate that a TVM with resonant responses near integer multiples of the signal frequency can amplify and up-convert an incident signal wave. By modulating the TVM with a pump frequency approximately twice the signal frequency we can overcome the power single-step conversion limits imposed by the Manley–Rowe relations. This is enabled by a cascaded process in which the pump first parametrically amplifies the signal wave and subsequently up-converts it to a sum-frequency that is approximately three times higher than the incident frequency. The proposed parametric system exhibits several characteristic features, including beating phenomena in the non-degenerate regime when operating close to the degenerate frequency ratio, as well as the requirement to maintain a precise frequency ratio and a fixed phase difference between the incident wave and the pump in the degenerate regime.

Our results provide the first demonstration of this effect in metasurfaces, and further improvements in the conversion efficiency are possible. Future work may include the design of meta-atoms that suppress radiation at the pump frequency, as well as the implementation of more efficient pump circuits. By minimising leakage of pump power into radiation at $f_p$, a larger fraction of the pump energy can be utilized for the conversion process. This would reduce the overall pump power required, which was a limiting factor in the experiments. Possible solutions include integrating a band-stop filter directly into the split-ring resonator structure, balanced or differential pumping schemes, or exploiting non-radiating modes at the pump frequency, all of which would suppress unwanted pump radiation and improve the effective energy transfer to the desired frequencies.

The saturation of the radiated power and the observed hysteresis are likely associated with the strongly nonlinear operation of the varactor diodes and the biasing network at high pump powers. In the experiments performed at a reduced dc bias voltage, strong ac modulation can drive the instantaneous diode voltage close to zero bias or even into forward bias during part of the modulation cycle. In this regime, the varactor no longer behaves as an ideal voltage-controlled capacitance and may exhibit self-biasing effects. In particular, the potential charging of capacitors in the dc supply line (used to provide additional protection of the dc source from ac currents) may effectively shift the dc bias across the diode. An increase in the effective reverse bias would reduce the average varactor capacitance, detune the resonant response, and modify the modulation efficiency. This self-biasing mechanism can also lead to hysteresis when the pump power is swept upward and downward. In addition, partial forward conduction and increased diode losses at large modulation amplitudes can reduce the quality factor, contributing to the saturation of the radiated power.

The integration of up-conversion and amplification within a single metasurface is highly promising for communication systems. Furthermore, as demonstrated, amplified conversion can be achieved at multiple sum-frequencies using a single signal frequency of the incident electromagnetic wave. This capability is particularly useful for designing frequency synthesizer systems in communication networks, enabling the utilization of multiple channels. Sum-frequency generation using parametric metasurfaces is also attractive for the implementation of compact terahertz radiation sources. Due to the widespread application of parametric amplification in ultrasensitive quantum computing readouts and receivers in radio astronomy and deep space communication, the proposed metasurfaces could significantly advance current technologies in this domain [31–33]. The obtained results can also be applied in the design of photonic time crystals [34,35].

## V. CONCLUSION

In this work, we investigated time-varying metasurfaces and characterized their performance for parametric frequency conversion, amplification, and wave generation. We demonstrated amplified up-conversion of free-space electromagnetic waves, enabled by a cascaded parametric process that arises when the metasurface is driven by a pump frequency close to twice the signal frequency and its resonances are tuned near integer multiples of the signal frequency. Under these conditions, the single-step power-conversion limits imposed by the Manley–Rowe relations can be surpassed. In addition, above a certain pump threshold, the metasurface enters a self-oscillation regime and operates as a generator of electromagnetic waves, predominantly at three-halves of the pump frequency. These results establish time-varying metasurfaces as a compact and efficient platform for frequency conversion, amplification, and wave generation in the microwave and terahertz regimes, with potential extension to higher frequencies using ultrafast modulation techniques and advanced material platforms.

*Contact author: fedor.kovalev@av.it.pt

## ACKNOWLEDGMENTS

This research was supported by the Australian Research Council Centre of Excellence for Transformative Meta-Optical Systems (Project ID CE200100010). S.M. and A.A. acknowledge funding by FCT – Fundação para a Ciência e a Tecnologia, I.P. (co-funded, when applicable, by EU funds) under the project UID/50008/2025 – Instituto de Telecomunicações, DOI:10.54499/UID/50008/2025. F.K. acknowledges support from the Robert and Helen Crompton Award.
The authors thank David Powell, Andrea Alù, Akshaj Arora, Puneet Garg, Zachary Fritts and Vitali Kozlov for useful discussions, as well as Michael Blacksell, Steve Marshall, Matthew Holding, José Carlos Pedro, Luís Cótimos Nunes and António Correia for assistance and guidance in preparing the experiment.

*Contact author: fedor.kovalev@av.it.pt

*Contact author: fedor.kovalev@av.it.pt

# Supplementary material for the paper "Amplified up-conversion of electromagnetic waves using time-varying metasurfaces"

Fedor Kovalev,[1,2,*] Stanislav Maslovski,[2] Abdelghafour Abraray,[2] and Ilya Shadrivov[1]

[1] ARC Centre of Excellence for Transformative Meta-Optical Systems (TMOS), Research School of Physics, The Australian National University, Canberra ACT 2601, Australia
[2] Instituto de Telecomunicações and Department of Electronics, Telecommunications and Informatics, University of Aveiro, Aveiro, 3810 – 193, Portugal

Figure S1 illustrates the schematic of the varactor diode MGV100-20 SPICE model, along with the pump and biasing circuit attached to the meta-atom in CST Studio. The SPICE model incorporates the parasitic capacitance and inductance of the 0805-02 package and the resistance of the varactor diode: $R_s =$ 0.3 Ohm, $C_p =$ 0.06 pF, $L_p =$ 0.4 nH. The varactor diode is modulated through a filtering circuit consisting of a passband LC filter at $f_p$ ($C_{ac}$ = 0.1 pF and $L_{ac}$ = 370 nH, with the parasitic capacitance $C_{pac}$ = 0.0685 pF) connected in series with an ac source of 50 Ω impedance, and a choke connected to a dc source ($L_{dc}$ = 500 nH with the parasitic capacitance $C_{pdc}$ = 0.063 pF). The passband filter at $f_p$ is designed for protecting the microwave generator from the dc and ac currents at $f_s$, $f_{sum1}$ and $f_{sum2}$. The circuits incorporate the parasitic capacitances of the inductors, which significantly affect the characteristics of the time-varying meta-atoms due to the small capacitance of the varactor diodes.

Figure S2 presents the meta-atom design consisting of two perpendicularly arranged printed circuit boards (PCB) based on the Rogers RO4350B substrates of thickness 0.51 mm, with dielectric permittivity $\varepsilon_r = 3.66$ and loss tangent $tan\delta = 0.0037$ used in CST Studio modeling. The resonator parameters are $R_1 = 21$ mm, $R_2 = 12$ mm, $w = 1.116$ mm, $g_1 = 1$ mm, $g_2 =$ 0.6 mm, $d = 0.543$ mm. The microstrip dimensions are $w = 1.116$ mm, $g_2 = 0.6$ mm, $g_3 =$ 0.45 mm, $w_{\mathrm{dc}} = 0.64$ mm, $l_{\mathrm{dc}} = 19.434$ mm, $a_{\mathrm{dc}} = 0.78$ mm, $l_{\mathrm{ac}} = 10.456$ mm, $l_{\mathrm{m1}} =$ 23.616 mm, $l_{\mathrm{m2}} = 6.508$ mm, $l_{\mathrm{m3}} = 2$ mm. The pad connecting the resonators to ground has the following size: $w = 1.116$ mm and $w_{\mathrm{p}} = 1.081$ mm, with via diameter of 0.6 mm. The ac microstrip width ($w$) was selected to realise a 50-Ω transmission line at $f_p$. Copper thickness is 35 μm throughout.

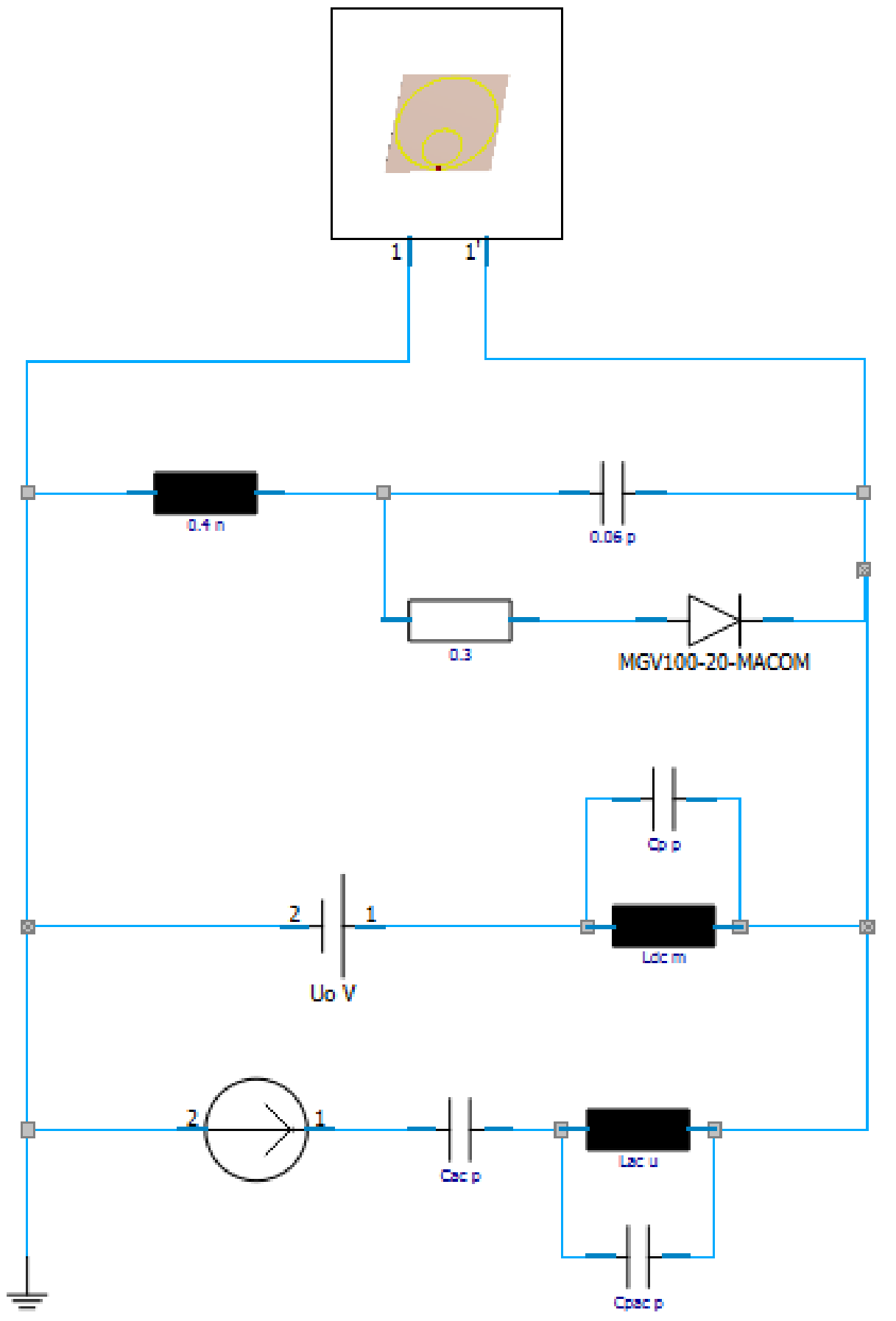

FIG S1. Schematic of the varactor diode MGV100-20 SPICE model, along with the pump and biasing circuit attached to the meta-atom in CST Studio.

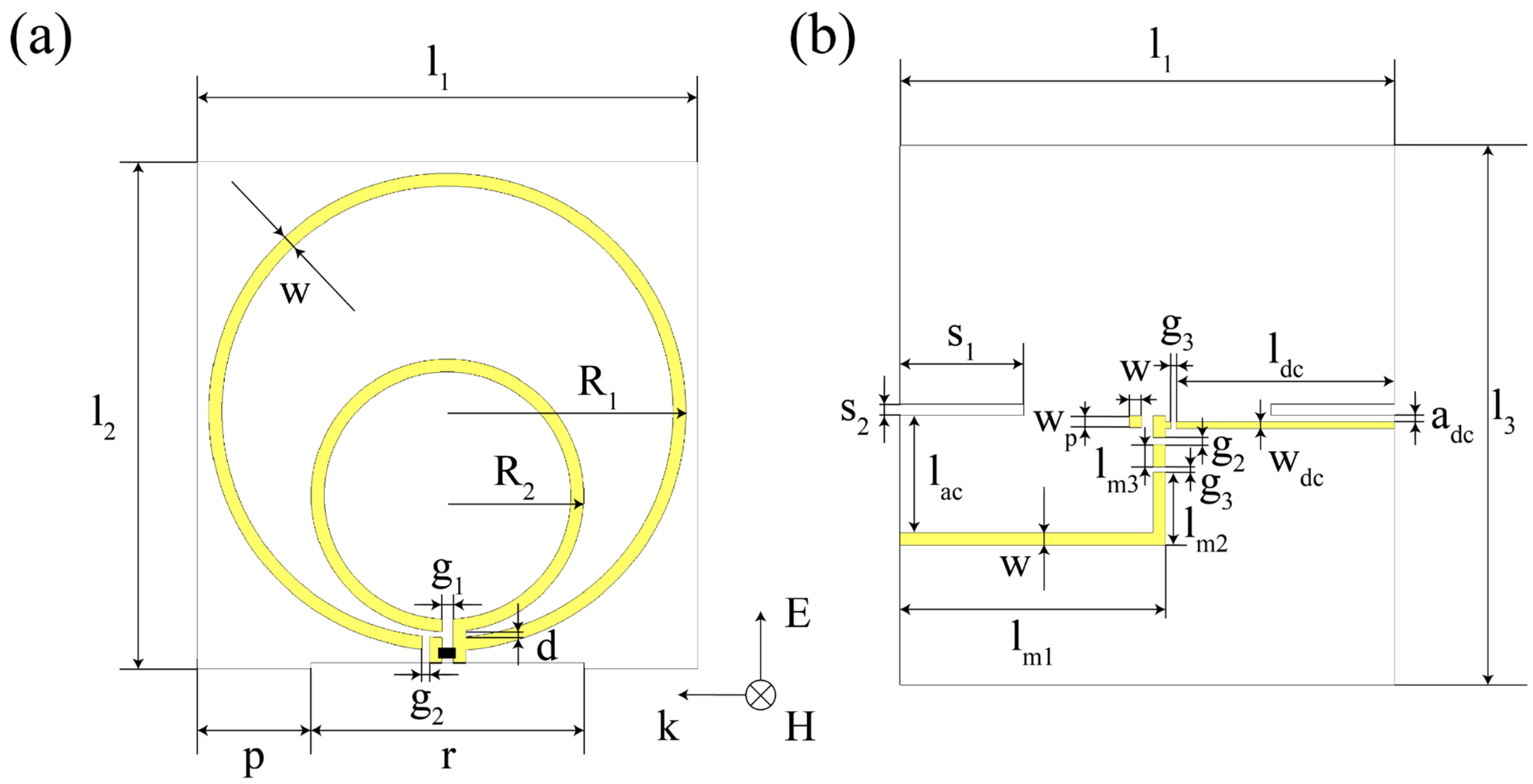

FIG S2. Meta-atom design: (a) vertical circuit board with two connected split-ring resonators and an embedded varactor diode (black square); (b) horizontal circuit board with the microstrips.

The circuit boards have dimensions $l_1 = 44$ mm, $l_2 = 44.51$ mm, $l_3 = 48$ mm. The horizontal board includes two slots ($s_1 = 11$ mm, $s_2 = 1.123$ mm) for inserting two protrusions of the vertical board ($p = 10$ mm) spaced by $r = 24$ mm. After soldering the lumped elements, the two circuit boards were glued together, and finally the pads of the horizontal board were soldered to the split-ring resonators. In our experiments, the ac microstrip line was connected to a coaxial cable using a standard coaxial connector, and the dc microstrip line was connected to a dc source using a regular dc cable.

In CST Studio simulations, we use discrete ports in the gaps and ends of the microstrip lines to simulate the corresponding circuit elements and sources. We conduct full-wave numerical simulations of the periodic structure using the finite-element method and transfer the 3D modeling results to the Schematic simulator in CST Studio to model the nonlinear lumped element circuit, which includes the varactor diode and the biasing network. S-parameters task was used for obtaining the reflection characteristics of the metasurface.

For obtaining the amplification and frequency conversion characteristics, we utilize the Spectral Lines task within the Schematic module of CST Studio to simulate the nonlinear circuit in the frequency domain. This approach leverages Harmonic Balance (HB) analysis, allowing us to investigate the interaction between the pump and signal within the designed circuit. The method accounts for frequency mixing up to a defined harmonic limit. To accurately capture the mixing effects, we consider 20 harmonics for the signal source and 10 harmonics for the pump source. In this study, the highest-order pump and signal harmonics considered in the simulations occur at nearly the same frequencies, ensuring that the harmonic balance analysis accurately captures their interaction.

Figures S3(a-d) demonstrate that the metasurface radiates more power at the sum frequencies in transmission than in reflection. While the reflected power at $f_s$ increases monotonically with pump amplitude, the transmitted power at $f_s$ initially decreases until $U_p \approx$ 8 V, after which it rises sharply with further increases. This behaviour is consistent with previously discussed parametric amplification phenomena in Ref. 25.

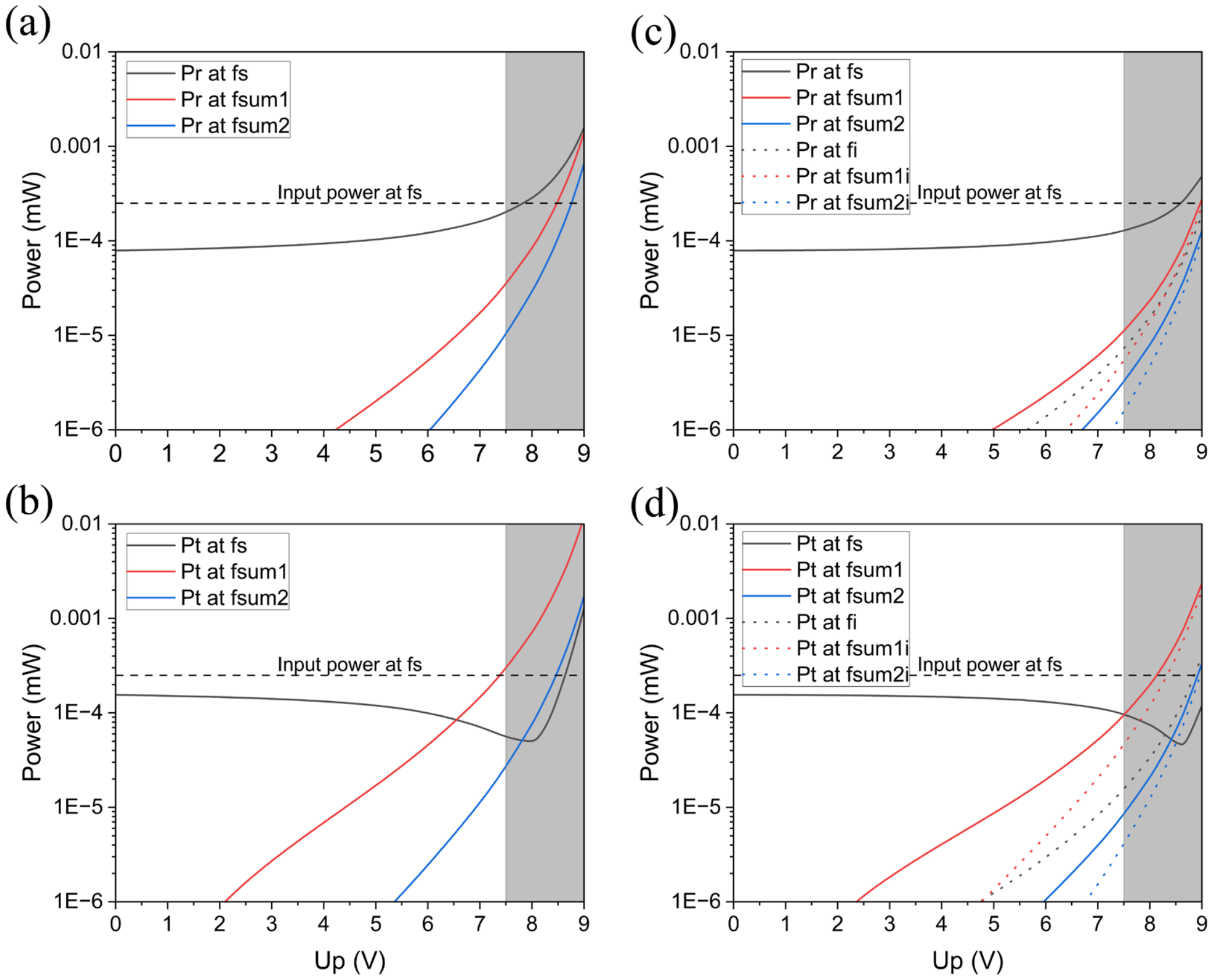

FIG S3. Radiated power versus pump amplitude for $U_0 = 10$ V (logarithmic scale). (a, b) Degenerate regime under phase-locking conditions with $f_s = 320$ MHz, $f_p = 640$ MHz, $f_{sum1} = 960$ MHz and $f_{sum2} = 1600$ MHz: (a) reflection and (b) transmission relative to the incident wave at $f_s$. (c, d) Non-degenerate regime with $f_s = 320$ MHz, $f_i = 319.9$ MHz, $f_p = 639.9$ MHz, $f_{sum1} = 959.9$ MHz, $f_{sum1i} = 959.8$ MHz, $f_{sum2} = 1599.8$ MHz and $f_{sum2i} = 1599.7$ MHz: (c) reflection and (d) transmission. The input power level at $f_s$ is indicated by the horizontal dashed line. The shaded region corresponds to $U_p > 7.5$ V, where the varactor model becomes inaccurate.

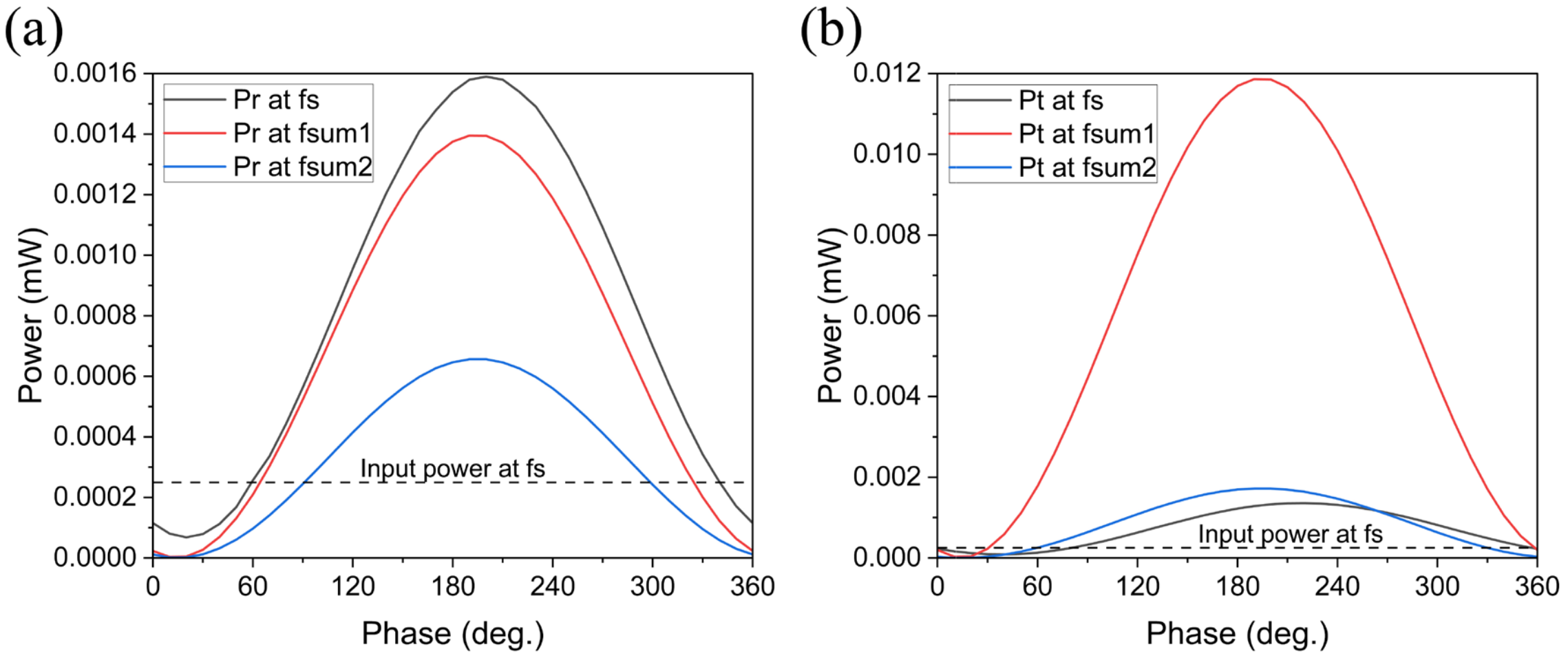

FIG S4. Radiated power at $f_s = 320$ MHz, $f_{sum1} = 960$ MHz and $f_{sum2} = 1600$ MHz versus pump phase at $f_p = 640$ MHz for $U_0 = 10$ V and $U_p = 9$ V in the degenerate regime: (a) reflection and (b) transmission relative to the incident wave at $f_s$.

Figure S4 shows the dependence of the radiated power at $f_s$ = 320 MHz, $f_{sum1}$ = 960 MHz and $f_{sum2}$ = 1600 MHz on the pump phase at $f_p$ = 640 MHz for $U_0$ = 10 V and $U_p$ = 9 V in the degenerate regime, in both reflection and transmission directions relative to the incident wave at $f_s$. The results in Figures 3(a) and S3(a, b) correspond to a pump phase of 190°, which yields the maximum radiated power at $f_{sum1}$. Figure S4 demonstrates that the radiated power at $f_s$, $f_{sum1}$ and $f_{sum2}$ follows a nearly sinusoidal dependence on the pump phase in the degenerate regime.

Figure S5 presents a schematic of the TEM cell used in the experiments. It consists of two parts: the ground and the stripline. The ground plane has dimensions of 684 mm × 305 mm. The stripline comprises three segments: a central part (228 mm × 252 mm) and two side parts (252 mm × 255.14 mm × 23 mm). The distance between the central stripline and the ground is 45 mm, while the shortest facets of the side parts are separated from the ground by 5 mm. The dimensions of the ground and the stripline were selected in a way to maintain approximately uniform characteristic impedance along the whole TEM cell.

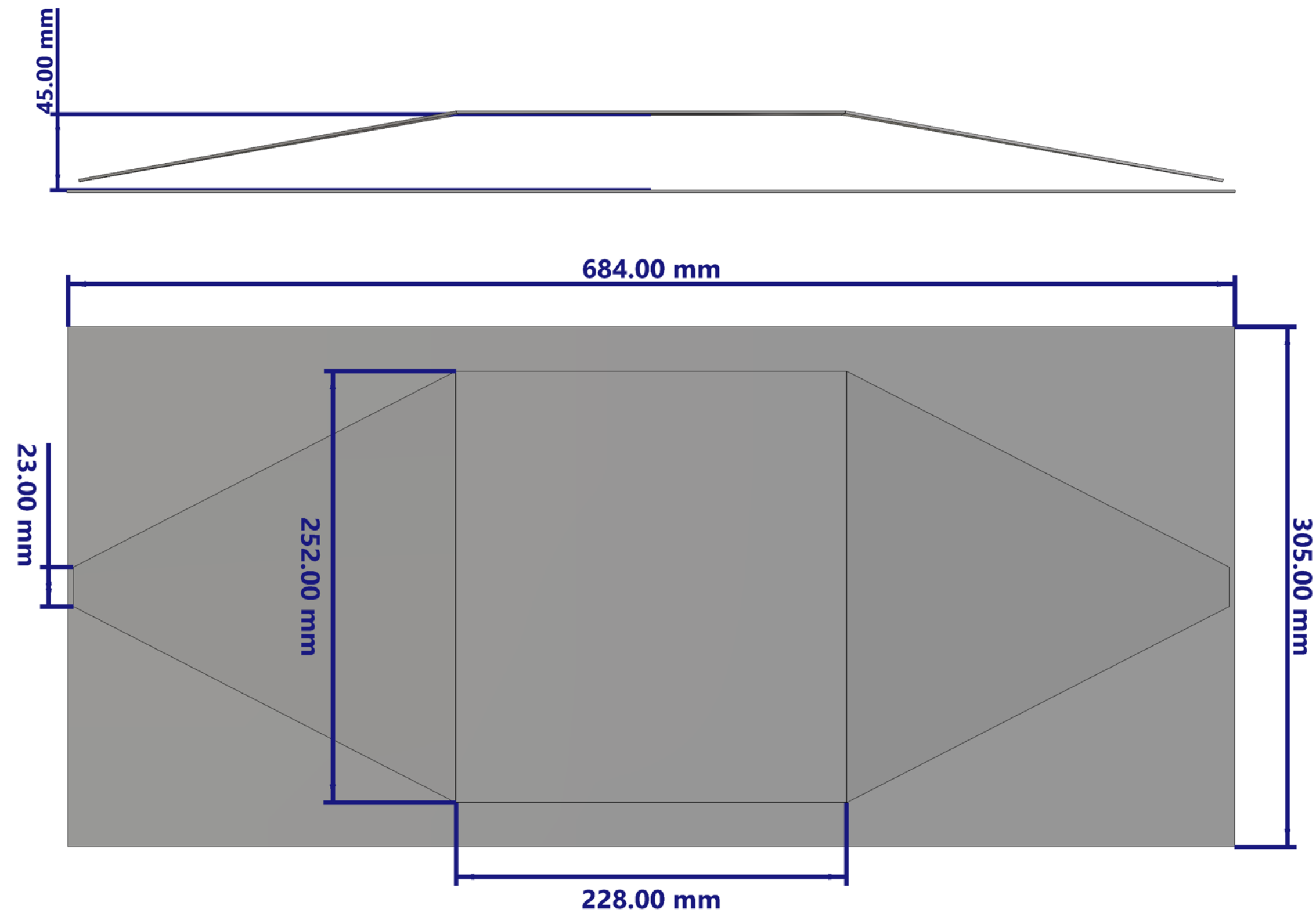

FIG S5. TEM cell and its dimensions.

Figure S6 illustrates the electric and magnetic fields in a plane perpendicular to the stripline and the ground plane when the TEM cell is excited at one port (shown as red cones) at 320 MHz. The fields are predominantly confined between the stripline and the ground plane, with the electric field oriented normal to these conductors and the magnetic field parallel to them. The stripline and the ground plane impose electric-wall boundary conditions, while the open lateral boundaries provide an approximate magnetic-wall symmetry. These symmetries are equivalent to the use of electric and magnetic mirror planes commonly employed to simulate infinite periodic structures. Consequently, when the meta-atom array is appropriately positioned with respect to these symmetry boundaries, the excitation conditions inside the TEM cell emulate those of an infinite periodic metasurface under plane-wave illumination.

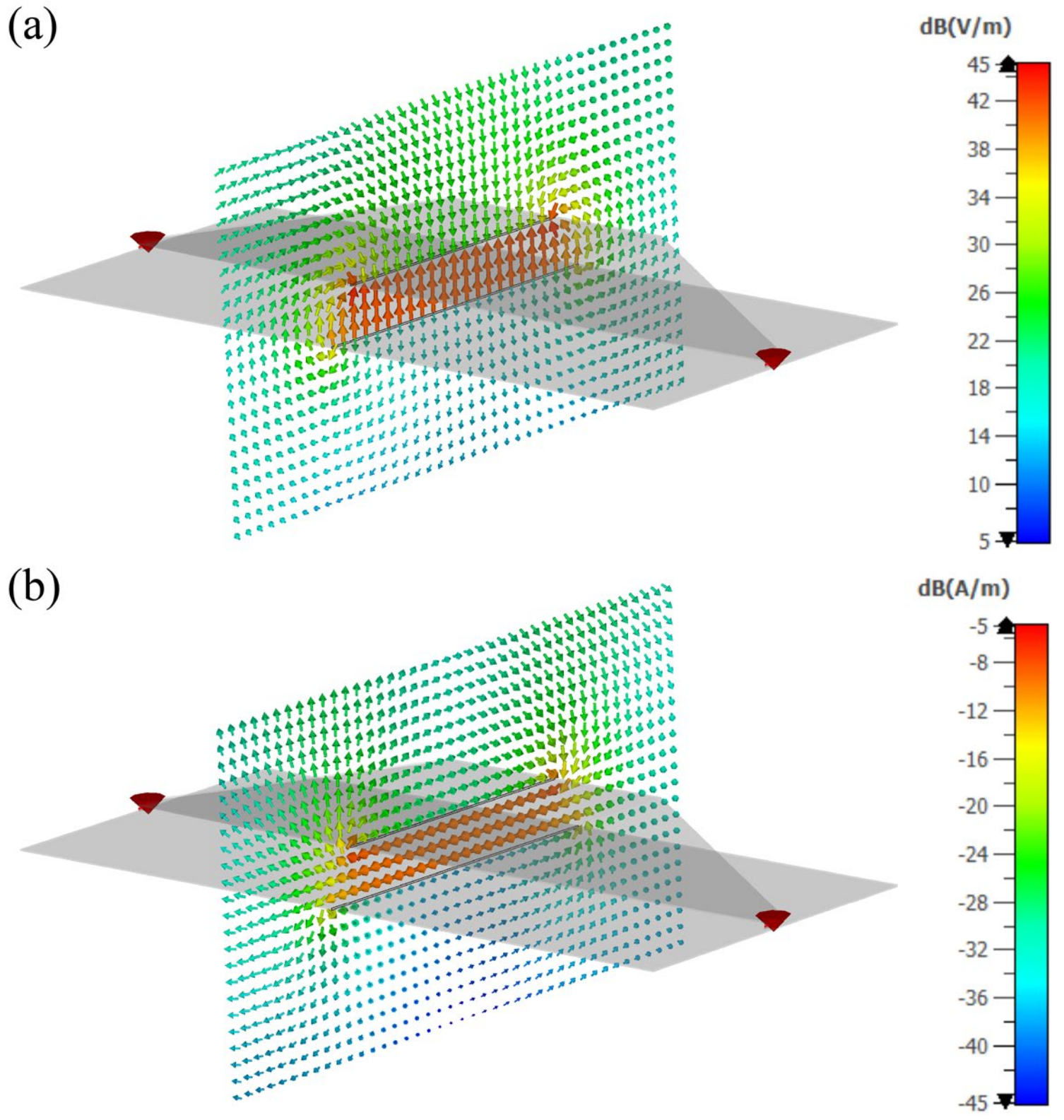

FIG S6. (a, b) Electric and magnetic field in the plane perpendicular to the TEM cell stripline and ground when excited at 320 MHz from one of its ports, shown as red cones.

Figure S7 shows the modeled and measured S-parameters of the TEM cell. The measurements were performed using a Keysight PNA-X N5242A vector network analyzer. The results confirm the expected performance: although the measured reflection (S22) is slightly higher at high frequencies and the transmission (S21) at 960 MHz is −4 dB compared to −2 dB in simulations, the reflection remains below −10 dB for frequencies up to 1350 MHz, which is sufficient to accurately measure the up-conversion characteristics at $f_{sum1}$.

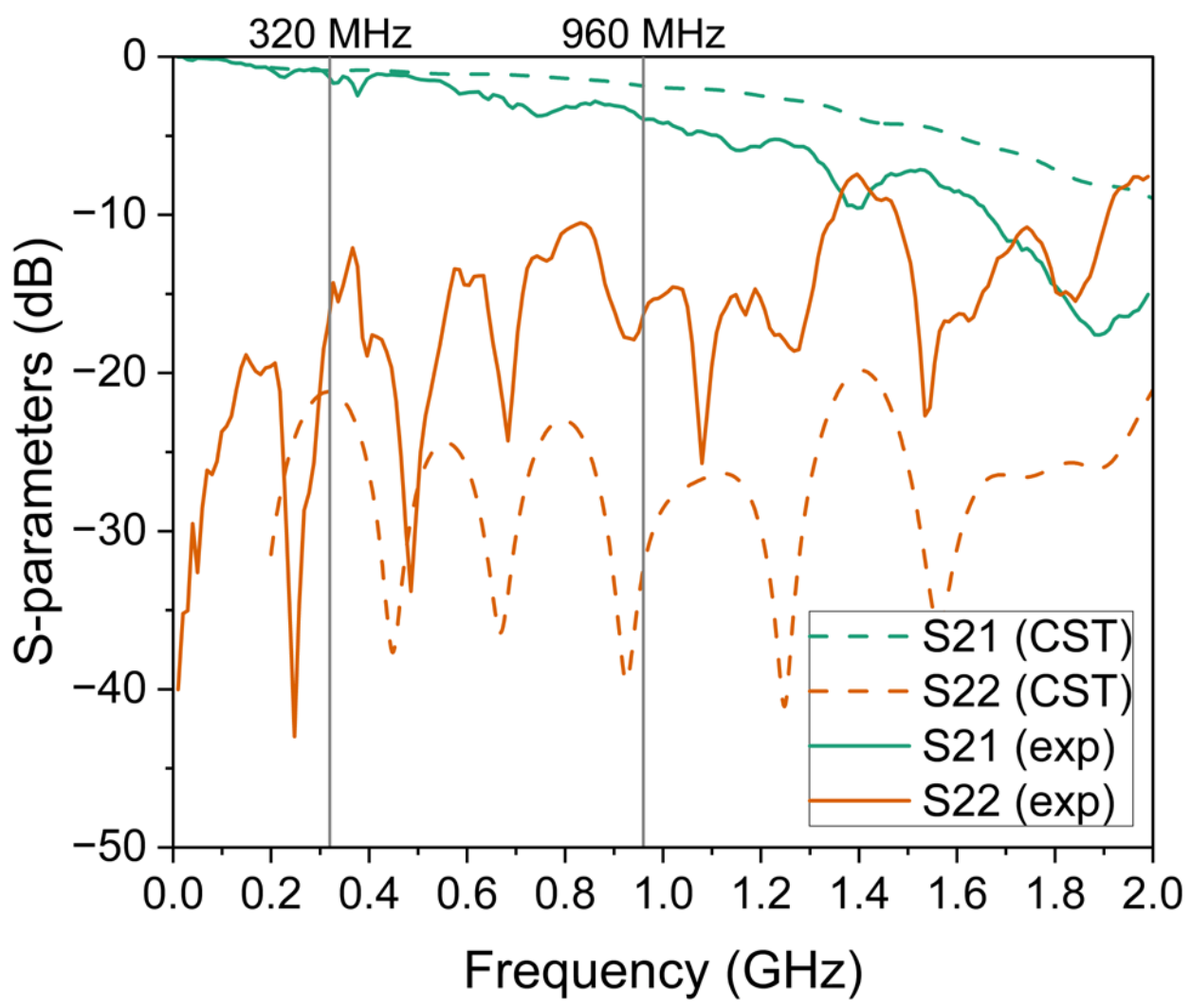

FIG S7. The simulated (dashed) and measured (solid) S-parameters of the TEM cell. The vertical lines at 320 and 960 MHz highlight the TEM cell performance around $f_s$ and $f_{sum1}$.

The developed TEM cell model was further used to obtain the scattering characteristics of seven meta-atoms excited at $f_s$ from one of its ports, as illustrated in Figure S8. For the numerical implementation, lumped ports loaded with equivalent inductance, capacitance, and resistance were employed to represent the meta-atom's lumped elements. The first lumped port in the SRR gap included $L_{dc} = 500$ nH and $C_{sum}$ (the sum of the varactor diode capacitance and parasitic capacitances; we used two values: $C_{max} = 1.2$ pF corresponding to $U_0 = 2$ V, $C_{min} = 0.4$ pF corresponding to $U_0 = 20$ V) in parallel. The second lumped port connected to the ac microstrip line incorporated $C_{ac} = 0.1$ pF, $L_{ac} = 370$ nH, and $R_g = 50$ Ω in series. The number of meta-atoms and their spacing were selected such that, in combination with the electric- and magnetic-wall symmetries of the TEM cell, the array emulates the response of an infinite periodic metasurface, ensuring in-phase excitation of all SRRs at $f_s$ and preserving the required tunability of the resonant responses.

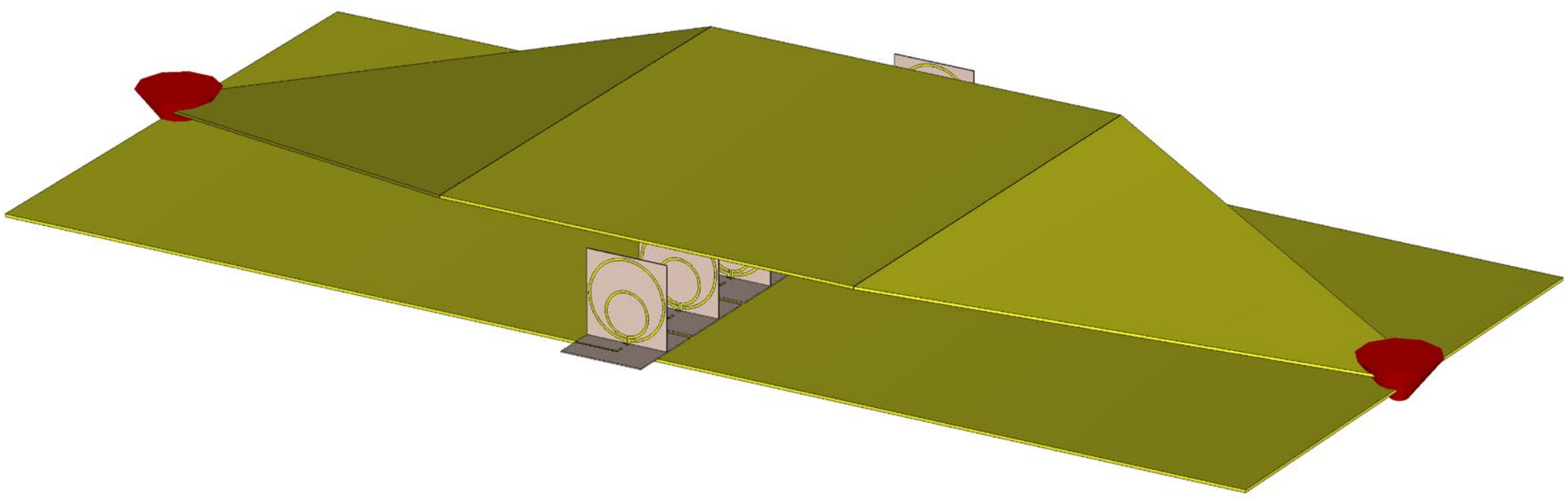

FIG S8. Model of the TEM cell with 7 meta-atoms placed on the ground plane.

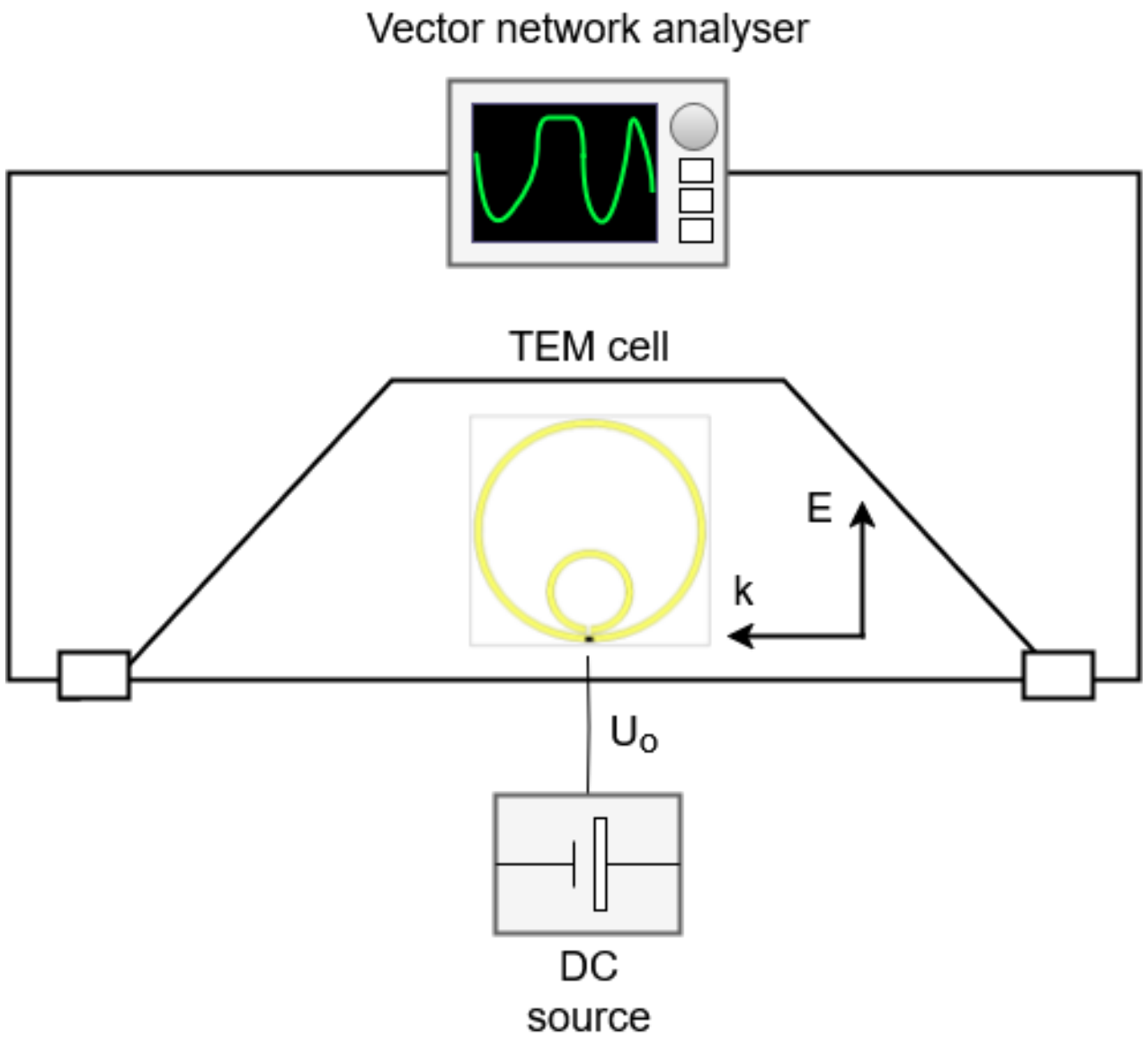

FIG S9. Diagram showing the S-parameters measurement scheme.

The S-parameters of the meta-atoms were measured using a Keysight PNA-X N5242A, as shown in Figure S9. Connections were made via vertical coaxial connectors mounted on the ground plane of the TEM cell, with their central conductors soldered to the stripline. For these measurements, the coaxial pump connectors were terminated with 50 Ω loads. Figure S9 also indicates the electric field direction and the wavevector of the excited TEM mode.

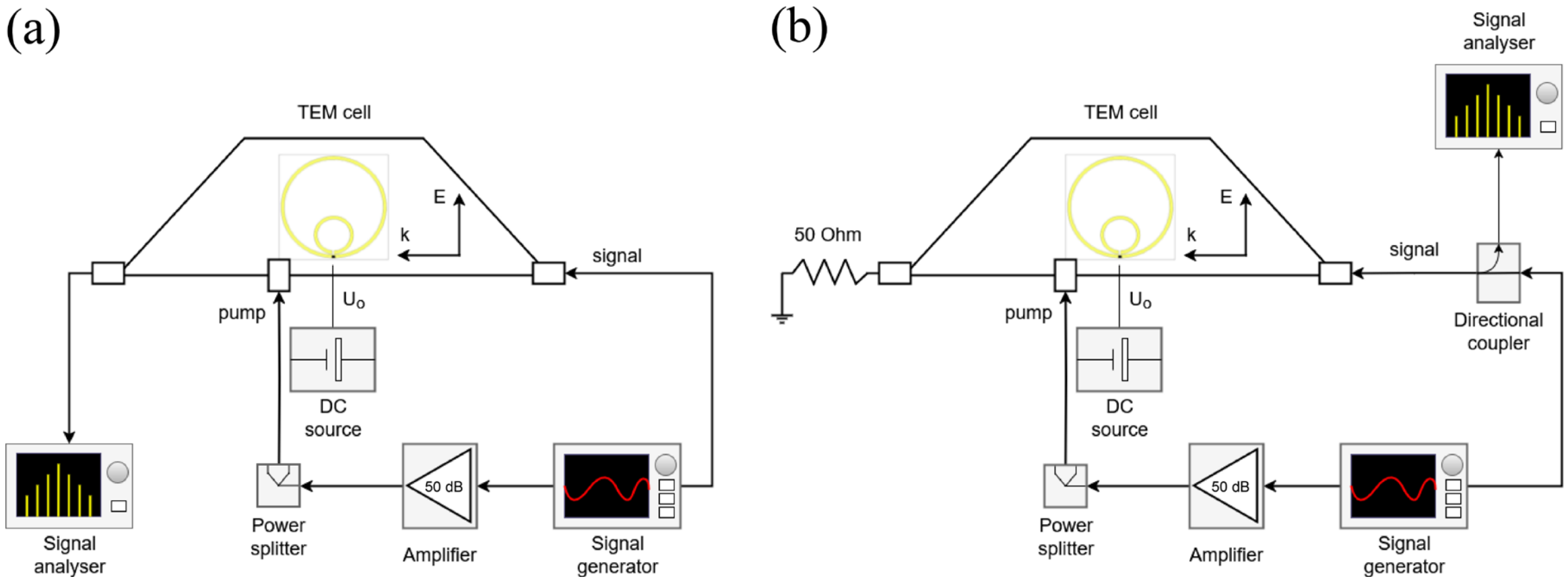

FIG S10. (a, b) Microwave experimental setup for measurements of up-conversion, amplification and generation characteristics of the time-varying meta-atoms in transmission and reflection.

Figure S10(a,b) presents the experimental setup used to measure up-conversion, amplification and generation of electromagnetic waves with the time-varying meta-atoms in the TEM cell, in transmission and reflection configurations, respectively. The system comprised a Rohde & Schwarz FSQ8 signal analyzer, a Rohde & Schwarz SMW200A two-port signal generator, a ZHL-30W-252-S+ power amplifier (providing 50 dB gain), a ZBSC-8-82+ power splitter, and an MX100QP dc source. A directional coupler (FDC-20-5-S+) was employed to measure the power of the radiated waves at the same port used to excite the incident wave, as it is shown in Figure S10(b). To enable this configuration, the opposite port was terminated with a 50 Ω load to absorb propagating waves. Since only one port could be monitored at a time, the measurements in reflection and transmission had to be performed separately.

Additionally, we analyze the phase dependence of the radiated power in the degenerate regime. Figure S11 shows the radiated power at $f_s = 313$ MHz, $f_{sum1} = 939$ MHz, and $f_{sum2} = 1565$ MHz in the directions of reflection and transmission as a function of the pump phase for $U_0 = 3$ V and a pump power of 32.7 dBm at $f_p = 626$ MHz. While the radiated power in reflection remains below the incident wave power (shown as a horizontal dashed line), the radiated power at $f_{sum1}$ in transmission can exceed this level when the pump phase is properly aligned with the signal phase. In addition, the minima and maxima of the radiated waves at $f_s$ are shifted differently relative to those at the sum-frequencies in both reflection and transmission. Although the experimental setup does not allow a direct comparison of reflection and transmission phase dependence (the directional coupler introduces an additional phase shift in reflection measurements), the results nevertheless confirm the dependence of the radiated waves on the pump phase in the degenerate regime and show good agreement with numerical modeling (see Figure S4).

(a)

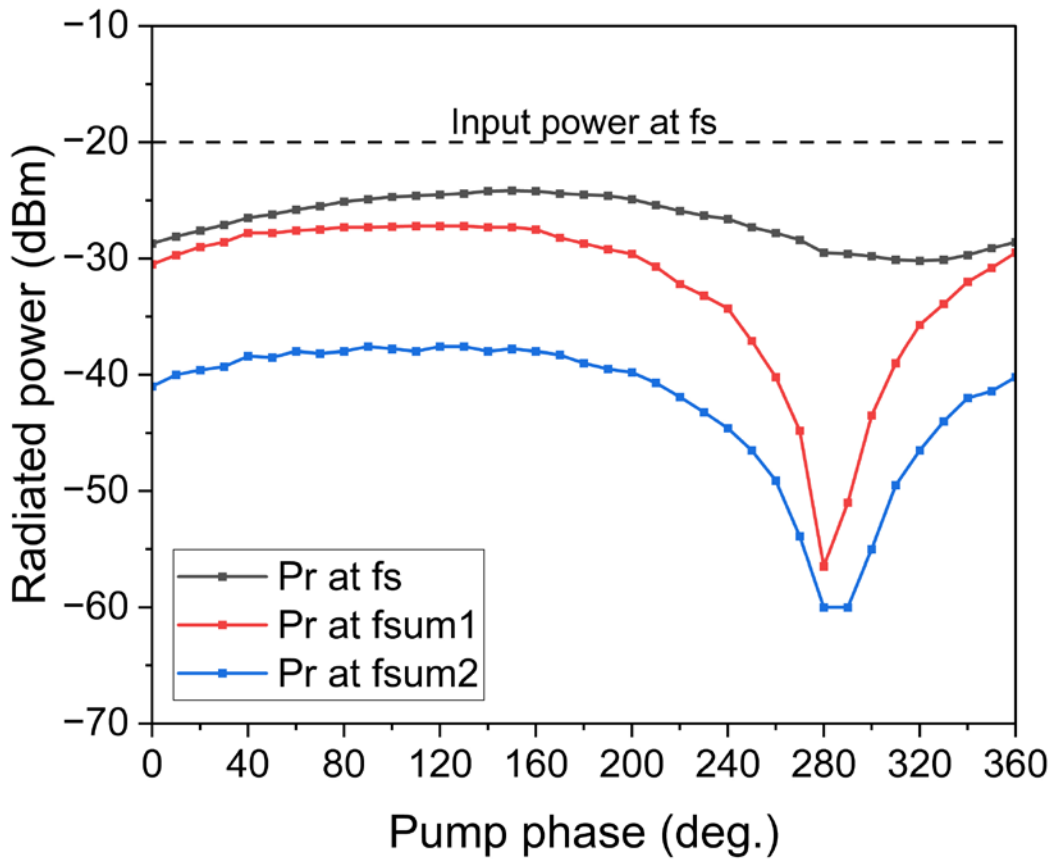

(b)

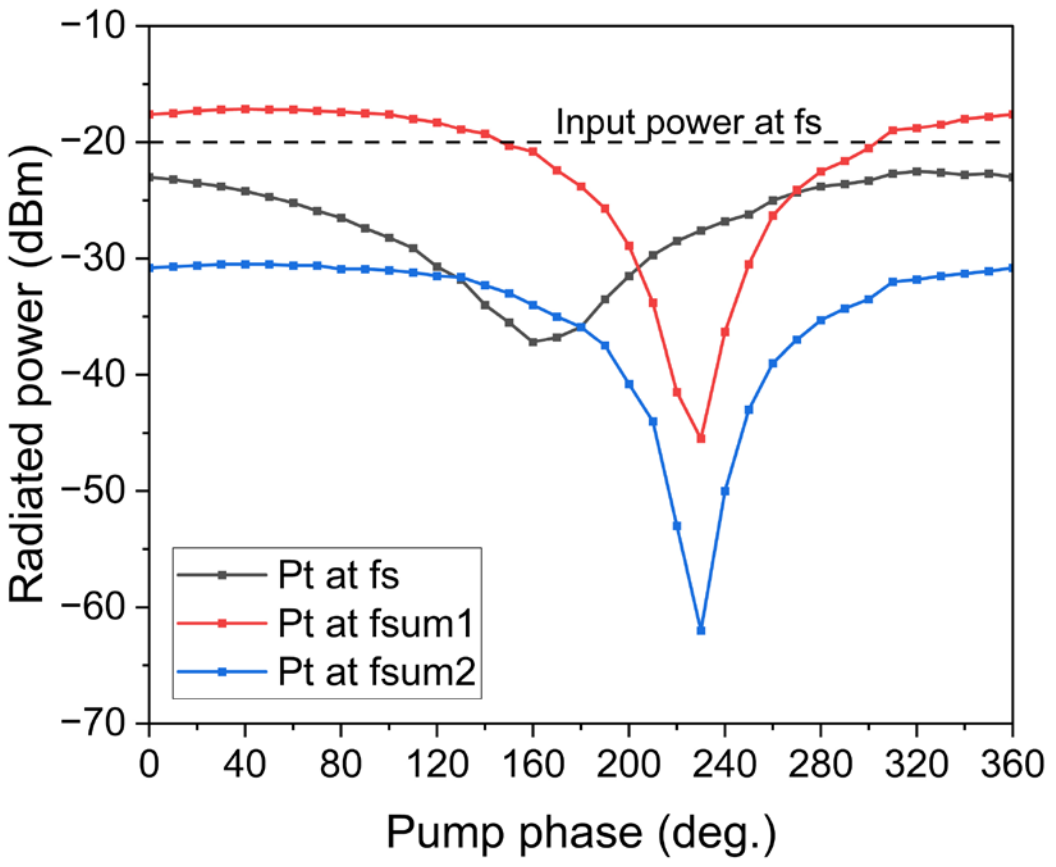

FIG S11. (a, b) Dependence of the radiated power at $f_s = 313$ MHz, $f_{sum1} = 939$ MHz, and $f_{sum2} = 1565$ MHz on the pump phase for $U_0 = 3$ V and a pump power of 32.7 dBm at $f_p = 626$ MHz in the degenerate regime. The horizontal dashed line indicates the input power level at $f_s$.

In this work, we aimed to design a metasurface analogue of a two-stage parametric amplifier, as schematically illustrated in Fig. S12. The proposed implementation is based on a time-varying meta-atom consisting of two connected split-ring resonators loaded with an embedded varactor diode. The effective resistances $R_{SRR1}$ and $R_{SRR2}$ include both radiation and dissipative losses of the corresponding split-ring resonators. $R_{SRR1}$ represents the input impedance associated with the incident wave $u(t)$, while $R_{SRR2}$ serves as the load of the two-stage parametric amplifier. $L_{SRR1}$ and $L_{SRR2}$ denote the inductances of the corresponding split-ring resonators, while $C_{SRR1}$ and $C_{SRR2}$ denote their effective capacitances. The time-dependent capacitance $C(t)$ represents the variable capacitance of the embedded varactor diode.

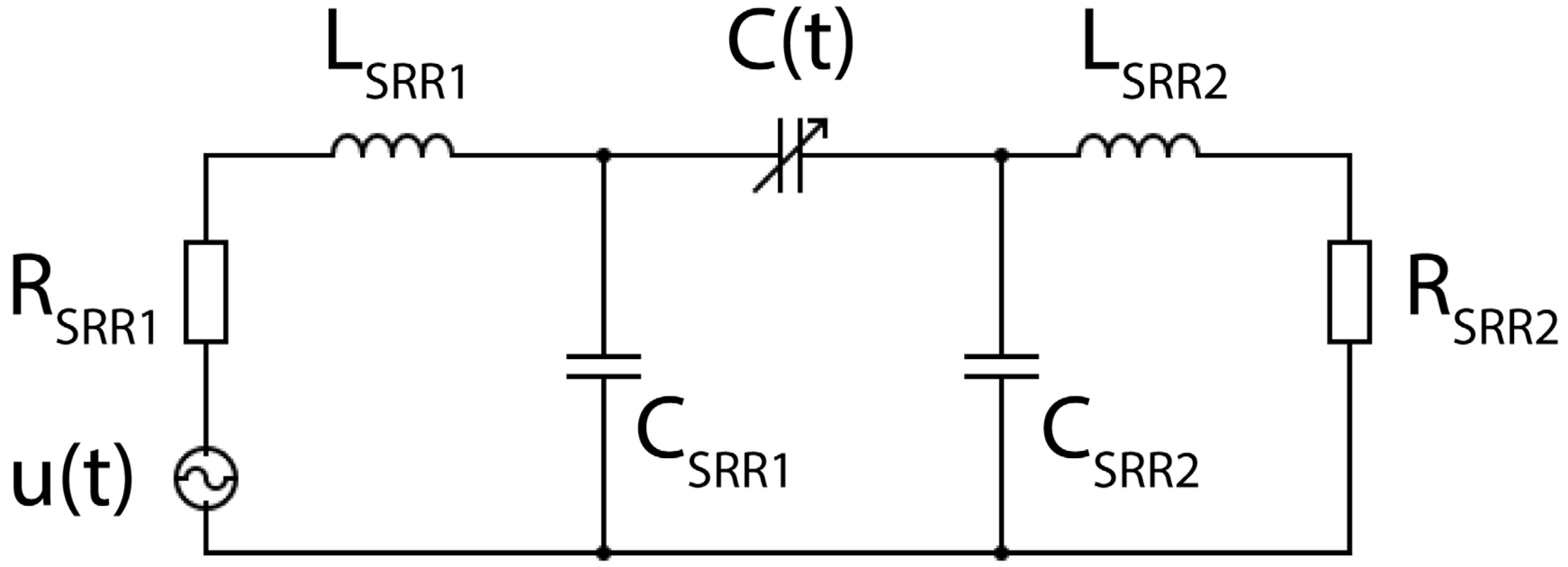

FIG S12. Equivalent circuit model of the time-varying meta-atom represented as a two-stage parametric amplifier.

Direct implementation of a classical two-stage parametric amplifier in a radiating metasurface is challenging, and it requires strong and efficient modulation of the relevant resonant modes. To better describe the experimentally implemented meta-atom, we therefore consider an equivalent circuit model that includes coupling between the two split-ring resonators. Figure S13 shows the equivalent circuit model of the time-varying meta-atom consisting of two connected split-ring resonators loaded with an embedded varactor diode. This model accounts for both inductive coupling, represented by the mutual inductance ($M$), and capacitive coupling, represented by the coupling capacitance ($C_c$), between the two resonators.

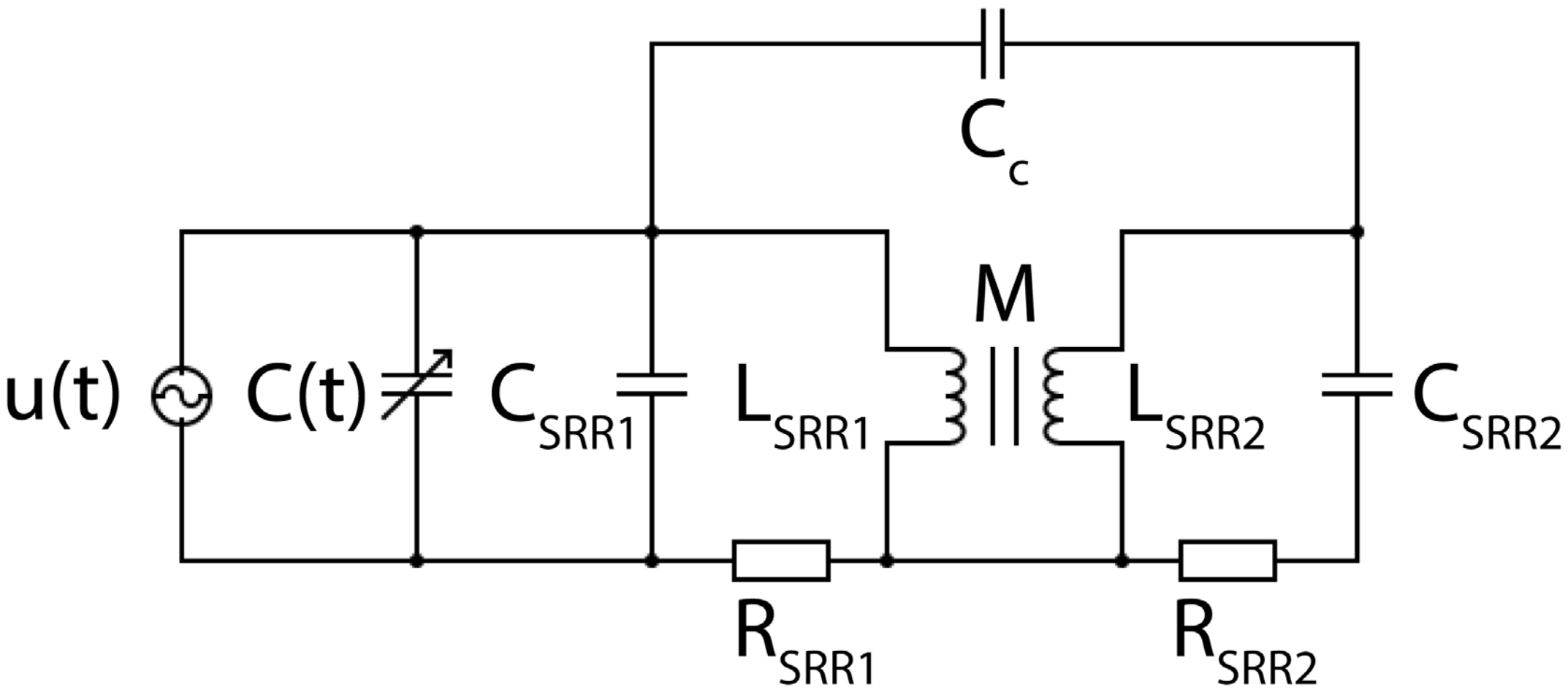

FIG S13. Equivalent circuit model of the experimentally implemented time-varying meta-atom, including inductive and capacitive coupling between the split-ring resonators.

Figure S14 shows a photograph of the TEM cell with the time-varying meta-atoms connected to the power splitter and placed inside an anechoic chamber. In the experiment, the modulation signal is supplied by a microwave signal generator followed by a power amplifier and is then divided among the meta-atoms. A pump power of 37 dBm measured at the output of the power amplifier corresponds to approximately 450 mW of modulation power at each ac pump coaxial input (26.5 dBm), before the LC passband filter connected to the corresponding varactor diode. This estimate accounts for the division of power among the output channels of the splitter. For this pump-power level, the measured radiated power at the pump frequency was approximately 12.6 dBm, corresponding to 18.4 mW. The exact modulation voltage across each varactor diode cannot be accurately determined from this value alone, because it depends on radiation losses at the pump frequency, the frequency-dependent response of the power splitter, cable and connector losses, the LC passband filter, impedance matching, reflections, diode-to-diode variations, and possible leakage through the dc bias line and other parasitic pathways. For this reason, the numerical simulations are presented as a function of the voltage directly applied to the varactor diode, whereas the experiments are presented as a function of the externally controlled pump power at the amplifier output. In practical implementations, pump radiation could be further reduced by improved filtering, balanced or differential pumping schemes, and meta-atom designs that suppress radiation at the pump frequency while preserving strong capacitance modulation.

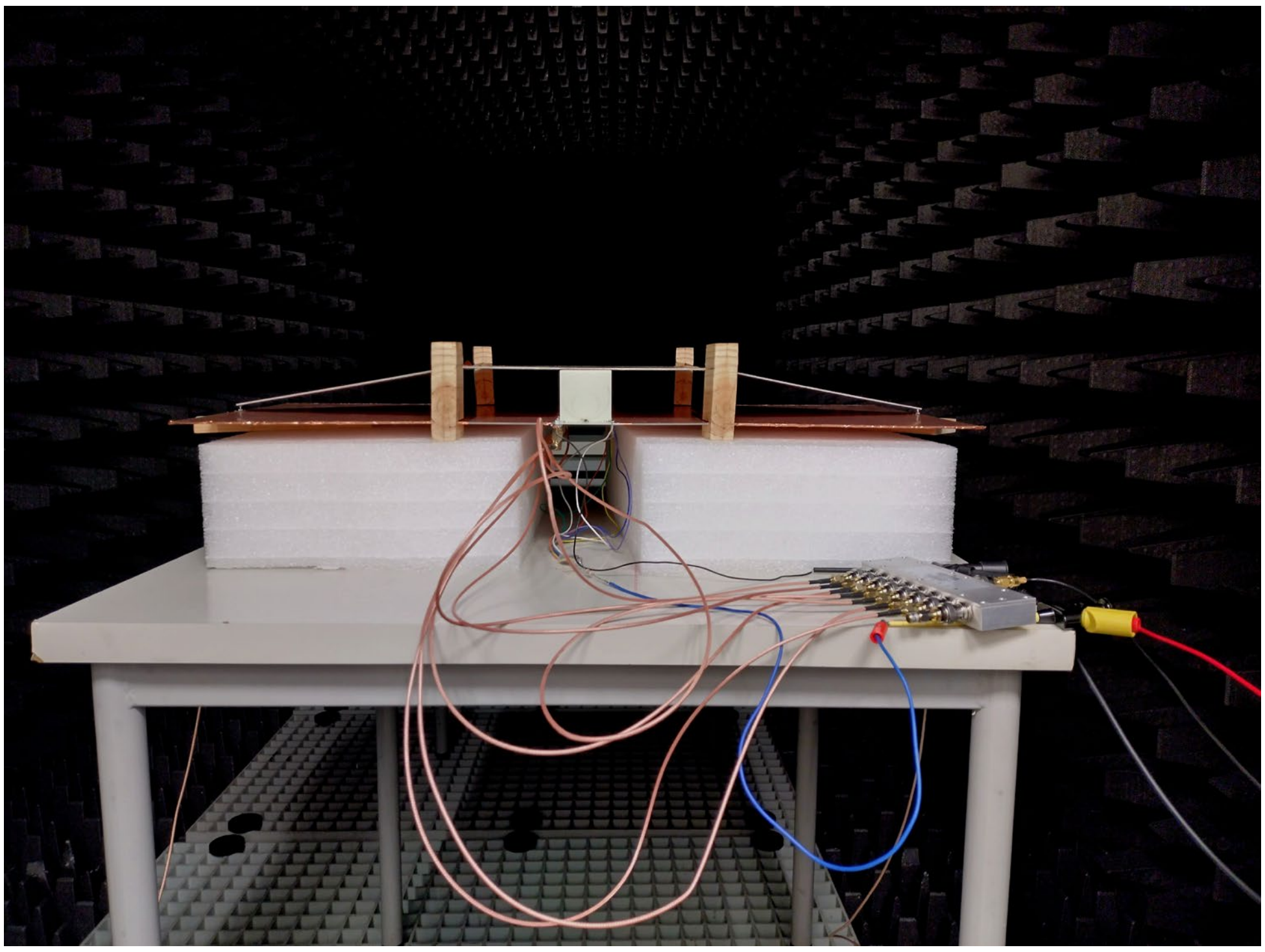

FIG S14. A photograph of the TEM cell and the time-varying meta-atoms connected to the power splitter and placed inside an anechoic chamber.